\documentclass[preprint,showpacs,preprintnumbers,amsmath,amssymb]{revtex4}

\usepackage{graphicx}
\usepackage{dcolumn}
\usepackage{bm}
\usepackage[]{amsmath,amssymb,epsfig,color}
\newcommand{\be}{\begin{equation}}
\newcommand{\ee}{\end{equation}}
\begin{document}


\title{Quantum chaos and its kinetic stage of evolution}

\author{L. Chotorlishvili$^{1,2}$, A. Ugulava$^{2}$}

%

 \affiliation{$1$ Institute for Physik, Universitat Augsburg, 86135 Augsburg,Germany}

\affiliation{$2$ Physics Department of the Tbilisi State
University,Chavchavadze av.3,~~0128, Tbilisi, Georgia}
\date{\today}
%

\begin{abstract}

Usually reason of irreversibility in open quantum-mechanical
system is interaction with a thermal bath, consisting form
infinite number of degrees of freedom. Irreversibility in the
system appears due to the averaging over all possible realizations
of the environment states. But, in case of open quantum-mechanical
system with few degrees of freedom situation is much more
complicated. Should one still expect irreversibility, if external
perturbation is just an adiabatic force without any random
features? Problem is not clear yet. This is main question we
address in this review paper. We prove that key point in the
formation of irreversibility in chaotic quantum-mechanical systems
with few degrees of freedom, is the complicated structure of
energy spectrum. We shall consider quantum mechanical-system with
parametrically dependent energy spectrum. In particular, we study
energy spectrum of the Mathieu-Schrodinger equation. Structure of
the spectrum is quite non-trivial, consists from the domains of
non-degenerated and degenerated stats, separated from each other
by branch points. Due to the modulation of the parameter, system
will perform transitions from one domain to other one. For
determination of eigenstates for each domain and transition
probabilities between them, we utilize methods of abstract
algebra. We shall show that peculiarity of parametrical dependence
of energy terms, leads to the formation of mixed state and to the
irreversibility, even for small number of levels involved into the
process. This last statement is important. Meaning is that, we are
going to investigate quantum chaos in essentially quantum domain.

In the second part of the paper, we will introduce concept of
random quantum phase approximation. Then along with the methods of
random matrix theory, we will use this assumption, for derivation
of muster equation in the formal and mathematically strict way.

Content of the paper is based on our previous studies. However, in
this review paper is included also some original material. This
part mainly concerns to the discussion about possible experimental
realization of theoretical concepts in the field of organic
chemistry.

$\bf keywords:$  Statistical Physics, Quantum Chaos, Nonlinear
Resonance, Open Quantum-Mechanical Systems.
\end{abstract}

\maketitle


\maketitle

$\bf{Introduction}$

The traditional notion of an area, where the laws of statistical
physics are effective, consists of the assumption that the number
of interacting particles is sufficiently large. However, a lot of
examples of nonlinear systems with a small number of degrees of
freedom, where chaotic motions occur, had become known by the end
of last century \cite{Sagdeev,Lichtenberg,Alligood}. A new stage
in the development of notions about chaos and its origin appeared
in the last two decades of the last century. It turned out that
classical Hamiltonian near the separatrix of topologically
different trajectories may experience a special kind of
instability. Because of this instability various dynamic
characteristics of the system randomly change with time. Such a
property of the system that performs random motion is called
dynamic stochasticity. Dynamic stochasticity is an internal
property of the system and is not associated with the action of
some a priory random forces.

Dynamic stochasticity appears due to the extreme sensitivity of
nonlinear system with respect to the slightly change of initial
conditions or systems parameters. On the other hand, even being
chaotic, dynamic is still reversible. Irreversibility occurs only
after averaging of the dynamic over small dispersion of initial
data. Note that averaging is not formal mathematical procedure. It
is essential from the physical point of view. Since dynamical
description loses its sense due to local instability of phase
trajectories.  However not the existence of initial error is
important but, what kind of consequences it has. In case of linear
system, this influence is negligible. So one can always assume
that initial data for linear system is defined with the absolute
accuracy. But in case of nonlinear systems, even small unavoidable
error should be taken into account. This leads to the necessity of
using concepts of statistical physics. As a result, analytical
description becomes much more complicated. All above mentioned was
concerned to the classical case. What is really happening in
quantum case? Should one still expect non-reversibility in quantum
case? The question is that as opposite to the classical case,
quantum equation of motion is linear. Of course, things are more
or less clear in case of open quantum systems interacting with the
thermostat. If so, then, irreversibility appears owing to the
averaging of systems dynamics over all possible realizations of
environment states.  Due to this, one can use standard formalism,
and from the Liouville-von Neumann equation, deduce irreversible
in time muster equation, for the reduced density matrix
\cite{Honggi}. But how does the irreversibility occur in quantum
systems with few degrees of freedom This is main question we
address in this review paper.

Usually quantum irreversibility is quantified by the fidelity.
Introduced by Peres \cite{Peres}, this concept works pretty well
and is especially convenient for the study of problems like
quantum chaotic billiards \cite{Jacquod}. On the other hand,
disadvantage is that obtaining of analytical results not always is
possible and large computational resources usually are needed.

In this paper, we offer alternative analytical method for the
study of problem of quantum irreversibility.  Our concept is based
on the features of the energy spectrum of chaotic quantum systems.
Peculiarity of energy spectrum of chaotic quantum systems is
well-known long ago \cite{Haake}. Random Matrix Theory (RMT)
presumes eigenvalues of chaotic systems to be randomly
distributed. Number of levels also should be quite large. In order
to deduce muster equation for time dependent chaotic system, we
will utilize methods of RMT in the second part of this paper.
However, what we want to discuses in the first part is completely
different. The key point is that, study of quantum chaos as a rule
is focused on the semi-classical domain. We mean not only
Gutzwiller's semi-classical path integration method \cite{
Stockman}, but also RMT. Since the RMT, in somehow implies
semi-classical limit. At least implicitly, due to the large number
of levels included into process.

In the first part of our paper, we shall consider chaotic
quantum-mechanical system with few levels included into process.
In spite of this, feature of the energy levels leads to the
irreversibility.

Paper is organized as follows:

In the first part we shall consider quantum mechanical-system with
parametrically dependent energy spectrum. This parametrical
dependence is quite non-trivial, contains domains of
non-degenerated and degenerated stats separated from each other by
branch points. Namely energy spectrum of our system is given in
terms of Mathieu characteristics. Due to the modulation in time of
the  parameter, system will perform transitions from one domain to
other one. For determination of eigenstates for each domain and
transition probabilities between them, we utilize methods of
abstract algebra. We shall show that peculiarity of parametrical
dependence of energy terms, leads to the formation of mixed state
and to the irreversibility, even for small number of levels
involved into the process. So this part may be considered as an
attempt to study quantum chaos in the essentially quantum domain.
This study is based on our previous papers
\cite{Ugulava,Chotorlishvili,Nickoladze,Gvarjaladze}. In addition,
in the present paper, we will discuss in details possible
experimental realizations and applications of the theoretical
concepts in the organic chemistry and polyatomic organic
molecules.

In the second part of the paper, we will introduce concept of
random quantum phase approximation. Then we will use this
assumption, for derivation of muster equation in the most formal
and mathematically strict way.

\section{Quantum Pendulum}

\subsection{Universal Hamiltonian}

Let us present the atom as a nonlinear oscillator under the action
of the variable monochromatic field. Then the Hamiltonian of the
system atom + field is of the form: \be H(x,p,t) =
H_{0}(x,p)+H_{NL}(x)+\varepsilon V(x,t),\ee where \be
H_{0}=1/2(\frac{p^{2}}{m}+\omega_{0}^{2}mx^{2}),~~H_{NL} = \beta
x^{3} + \gamma x^{4}+ \ldots,\ee \be V(x,t)=V_{0}x\cos \Omega
t,~\varepsilon V_0=\frac{e}{m}f_0,~\varepsilon\ll1.\ee Here $x$
and $p$ are coordinate and the impulse of the particle (electron),
$\omega_0$ is the  frequency of oscillations, $\beta$ and $\gamma$
are coefficients of nonlinearity, $m$ and $e$ are the mass and
charge of the particle, $f_0$ is the amplitude of the variable
field. Having made passage to the variables of action-angle
$I,\theta$ with the help of transformation $x=(2I/m\omega
_0)^{1/2}\cos\theta,p=-(2Im\omega _0)^{1/2}\sin\theta,$ supposing
resonance condition $\Omega\approx\omega _0$ is realized and
averaging with respect to the fast phase $\theta$, one can
obtain:$$H=H_0(I)+\varepsilon V(I)\cos\varphi,$$ where \be
H_0(I)=\omega_0I+H_{NL},~~H_{NL}(I)=\frac{3\pi}{4}(\frac{I}{m\omega^2
_0})^2\gamma \ee \be \varphi=\theta-\omega t,~~\varepsilon
V(I)=\sqrt{I/m\omega _0}V_0.\ee The role of nonlinear frequency
plays $(dH_0/dI)=\omega_0+\omega _{NL}(I)$ where $\omega
_{NL}(I)=(3\pi/2)(I\gamma/m\omega^2 _0)$. Now suppose that
nonlinear resonance condition $\omega_0+\omega _{NL}(I_0)=\Omega$
is fulfilled for action $I=I_0$. It is easy to show
\cite{Sagdeev}, that for a small deviation of action from the
resonance value $\Delta I\equiv I-I_0~(\Delta I\ll I_0)$ after the
power series expansion, if the condition of moderate nonlinearity
is just $\mu\ll1/\varepsilon$, where $$\mu\equiv\omega
_{NL}(I/\omega _{NL}(I))|_{I=I_0},~~\omega _{NL}=(d\omega
_{NL}(I)/dI)|_{I=I_0},$$ for Hamiltonian we obtain: \be H =
\frac{\omega_{NL}}{2}(\Delta I)^2+\varepsilon V(I_0)\cos
\varphi.\ee As usual (6) is called universal Hamiltonian. Let us
notice that Hamiltonian $H$ is the Hamiltonian of pendulum, where
$I/\omega _{NL}$ plays the role of mass, $\Delta I$ plays the role
of the pulse and $V(I_0)$ plays the role of potential energy. If
in (6) $\Delta I$ is substituted by the appropriate operator
$\Delta I\rightarrow -i\hbar\partial/\partial\varphi$, one can
obtain the universal Hamiltonian in the quantum form \be
H=-\frac{\hbar^2\omega'}{2}\frac{\partial^2}{\partial\varphi^2}+V\cos\varphi.
\ee With the help of (7) it is possible to explore quantum
properties of motion for the nonlinear resonance.

Having written the stationary Schrodinger equations \be
\hat{H}\psi_n=E_n\psi_n\ee for the Hamiltonian (7), we get \be
\frac{\partial^2\psi_n}{\partial\varphi^2}+(E_n-V(l,\varphi))\psi_n=0,
\ee $$V(l,\varphi)=2l\cos2\varphi$$ where the dimensionless
quantities are introduced \be
E_n\rightarrow\frac{8E_n}{\hbar^2\omega'},~l\rightarrow\frac{4V}{\hbar^2\omega'}\ee
and the replacement $\varphi\rightarrow 2\varphi$ is done.

\begin{figure}[t]
  \centering
  \includegraphics[width=12cm]{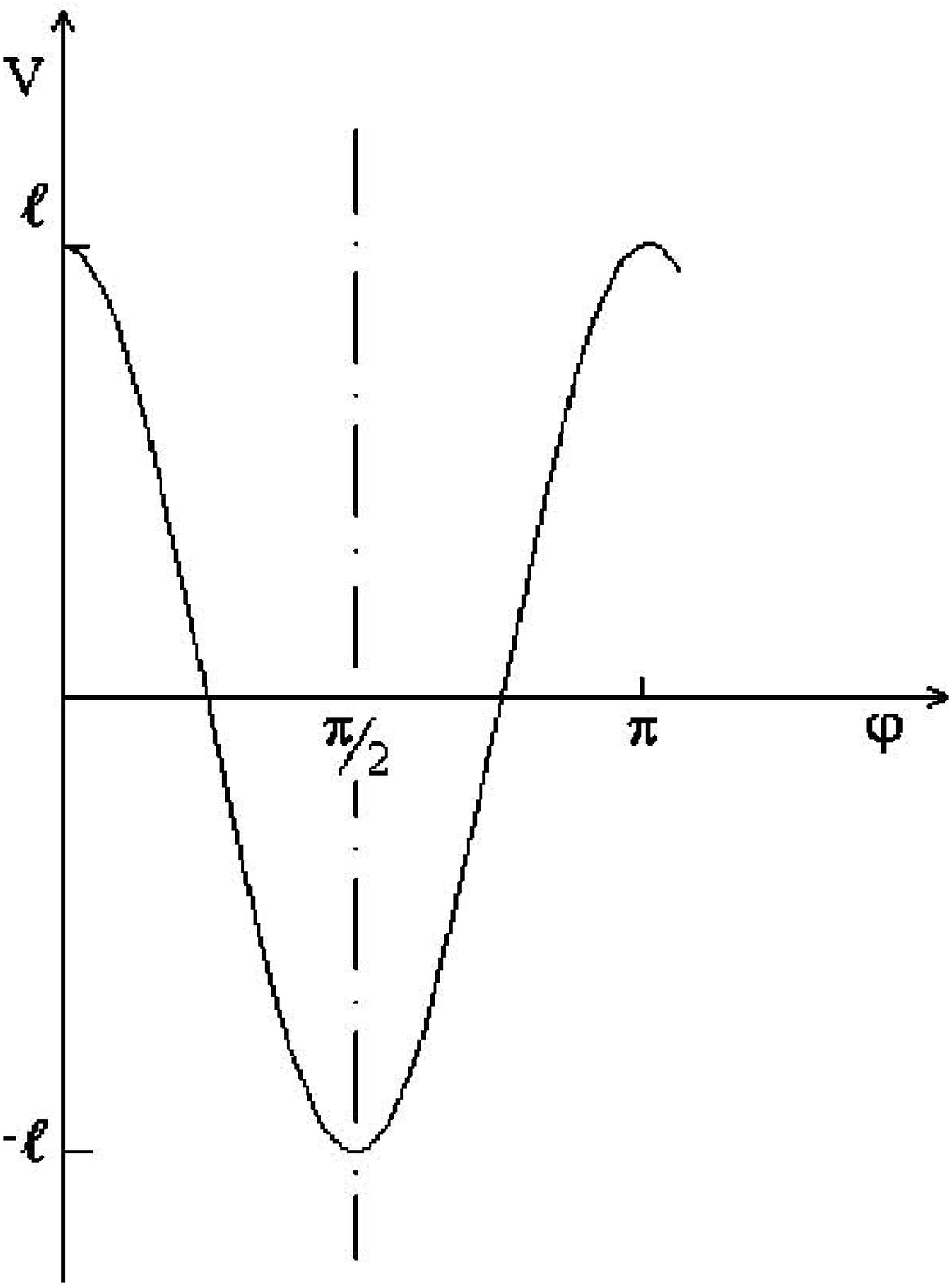}
  \caption{The dependence of the energy of interaction $V$ from the phase.}\label{fig:Fig.1}
\end{figure}

The interaction has the following properties of the symmetry: $
1.V(\varphi)=V(-\varphi),~2.V(\varphi)=V(\pi+\varphi),~3.V(\frac{\pi}{2}+\varphi)=V(\frac{\pi}{2}-\varphi).$
G.M.Zaslavsky and G.P.Berman [8] were the first who considered the
equation of the Mathieu-Schrodinger for the quantum description of
the nonlinear resonance in the approximation of moderate
nonlinearity. They studied the case of quasi-classical
approximation $\Delta I\gg\hbar$ for both variables $I_0$ and
$\Delta I$. In this review we investigate the equation of the
Mathieu - Schrodinger in essentially quantum area.

\subsection{Periodical solution of Mathieu - Schrodinger
equations}

We content ourselves with only even and odd solutions with respect
to $\varphi$ of equation of the Mathieu - Schrodinger (9). Those
solutions have $n$ zeroes in the interval $0\leq\varphi\leq\pi$.
Eigenfunctions $\psi_n$ can be recorded with the help of the
Mathieu functions [9,10]: even $ce_n(\varphi,l)$ and odd
$se_n(\varphi,l)$. Appropriate eigenvalues are usually designated
by $a_n(l)$ and $b(l)$. For simplicity below sometimes we omit the
argument $l$ and write $ce_n(\varphi),~se_n(\varphi),~a_n,~b_n.$
Mathieu functions are eigenfunctions of the problem of
Sturm-Liouville for the equation (9) for the boundary conditions
$$ \psi(0)=\psi(\pi)=0,~~~for~~~ se_n(\varphi)$$
\be
\frac{d\psi}{d\varphi}(0)=\frac{d\psi}{d\varphi}(\pi)=0,~~~for~~~
ce_n(\varphi).\ee From the general theory of the Sturm-Liouville
it follows, that for arbitrary $n=1,2,\dots$ there exists
eigenfunction $se_n(\varphi)$ and for each $n= 0,1,2,\dots$
determined eigenfunction $ce_n(\varphi)$. The definition of the
Mathieu function must be supplemented with the choice of the
arbitrary constant so that the conditions were fulfilled:
$$ce_n(0,l)>0,~~~\frac{1}{\pi}\int_0^{2\pi}ce_n^2(\varphi,l)d\varphi=1,$$
\be
\frac{dse_n}{d\varphi}(0,l)>0,~~~\frac{1}{\pi}\int_0^{2\pi}se_n^2(\varphi,l)d\varphi=1.\ee
If $\psi(\varphi)\equiv G(\varphi)$ means either $ce_n(\varphi)$
or $se_n(\varphi)$, then $G(\varphi)$ and $G(\pi-\varphi)$ satisfy
the same equation (9) and the same boundary conditions (11).
Therefore these functions differ from each other only by the
constants. Hence, $G(\varphi)$ is even or odd function with
respect to $\pi/2-\varphi$. Taking this into account, two
functions (11) break up into four Mathieu functions:
$$\psi(0)=\psi(\pi/2)=0,~~G(\varphi)=se_{2m+2}(\varphi),~~phase~~\pi,$$
$$\psi(0)=\frac{d\psi}{d\varphi}(\pi/2)=0,~~G(\varphi)=se_{2m+1}(\varphi),~~phase~~2\pi,$$
$$\frac{d\psi}{d\varphi}(0)=\psi(\pi/2)=0,~~G(\varphi)=ce_{2m+1}(\varphi),~~phase~~2\pi,$$
\be\frac{d\psi}{d\varphi}(0)=\frac{d\psi}{d\varphi}(\pi/2)=0,~~G(\varphi)=ce_{2m}(\varphi),~~phase~~\pi.
\ee

For arbitrary $m = 0,1,2,\dots$ there is one eigenfunction for
each of four boundary conditions and $m$ equal to the number of
zeroes in the interval $0<\varphi<\pi/2$. The functions (13)
represent a complete system of eigenfunctions of the equation (9).

\subsection{Symmetries of the equations of Mathieu-Schrodinger}

The properties of symmetry of the Mathieu-Schrodinger equation can
be presented in Table 1 \cite{Bateman}.

As is known, group theory makes it possible to find important
consequences, following from the symmetry of the object under
study. Below, with the aid of group theory we will establish the
presence (or absence) of degeneracy in the eigen spectrum of the
Mathieu-Schrodinger equation and a form of corresponding wave
functions By immediate check it is easy to convince, that four
elements of transformation
$$G(\varphi\rightarrow-\varphi)=a,~~~G(\varphi\rightarrow \pi-\varphi)=b,$$
$$G(\varphi\rightarrow \pi+\varphi)=c,~~~G(\varphi\rightarrow \varphi)=e$$

\begin{table}[t]
  \centering
 \includegraphics[width=16cm]{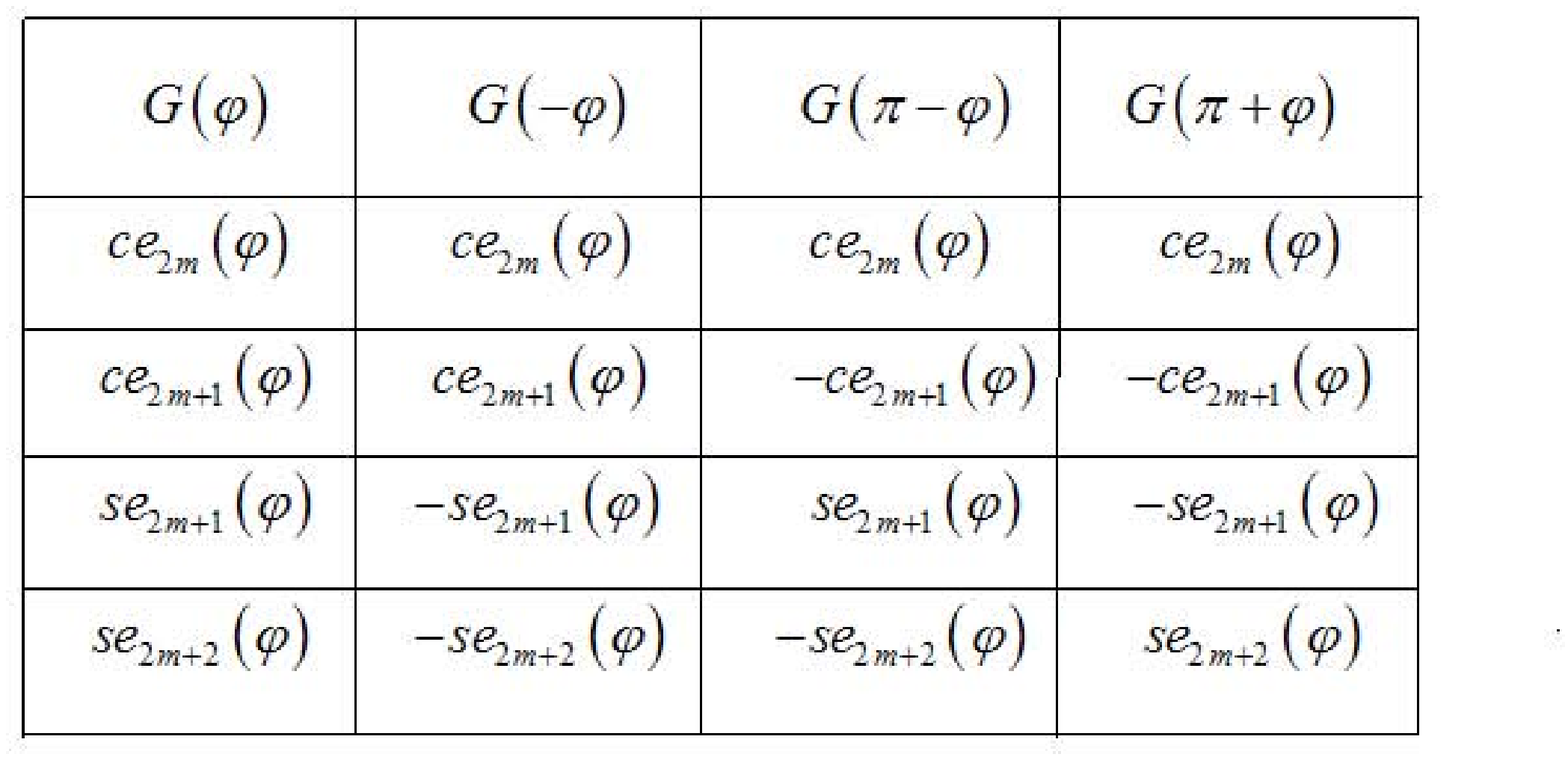}
  \caption{Relations of a symmetry for the Mathieu function.}\label{tab:Tab.1}
\end{table}
are forming a group. For this it is enough to test the realization
of the following relations:
$$a^2=b^2=c^2=e,$$ \be ab=c,~~ac=b,~~bc=a.\ee
The group contains three elements $a,b,c$ of the second order and
unity element $e$. The group $G$ is isomorphic to the well-known
quadruple group of the Klein \cite{Hamermesh,Landau}. This group
is known in the group theory by the applications to the quantum
mechanics (designated as $V$). All the elements of the group
commute. This assertion can be easily checked taking into account
group operations (14). So, the symmetry group of the Mathieu
function $G$ is the Abelian group and consequently has only
one-dimensional indecomposable representations.

The presence of only one-dimensional representations of the
symmetry group, describing the considered problem, hints on the
absence of degeneration in the energy spectrum. So, we conclude,
that the eigenvalues of the equation of the Mathieu - Schrodinger
(9) are non-degenerated, and the eigenfunctions are the Mathieu
functions (13). However we shall remind, that both the energy
terms $a_n,b_n$, and the Mathieu functions depend on the parameter
$l$. At the variation of $l$ in the system can appear symmetry
higher, than assigned in Table 1, that might lead to the
degeneration of levels.

In order to make more obvious the isomorphism of the symmetry
group $G(\varphi)$ of Mathieu-Schrodinger equation with quadruple
Klein group $V$, let us consider the plane of rotation of phase
$\varphi$.

\begin{figure}[t]
  \centering
  \includegraphics[width=12cm]{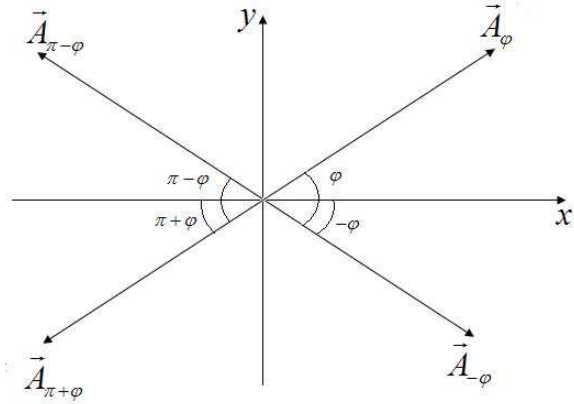}
  \caption{Four vectors that are transformed by the elements of quadruple group $V$.}\label{fig:Fig.2}
\end{figure}

The orientation of vector  $A_{\pi-\varphi}$, presented in Fig.2,
is obtained from $A_\varphi$ by means of mirror reflection
relative to the plane passing through the axis $0y$ perpendicular
to the figure plane $\sigma_y$. The orientation of vector
$A_{\pi+\varphi}$ is obtained by means of rotation by angle $\pi$
about an axis, passing perpendicular to the figure through the
origin of the coordinate system  $C_2$ (axis of rotation of the
second order). The symmetry elements $\sigma_x,~\sigma_y$ and
together with the unit element form quadruple group
$V:~e,~\sigma_x,~\sigma_y,~C_2$. Now it is possible to bring the
elements of two group to one-to-one correspondence:
$G(-\varphi)\rightarrow\sigma_x,~G(\pi-\varphi)\rightarrow\sigma_y,~
G(\pi+\varphi)\rightarrow C_2,~e\rightarrow e$ that proves the
isomorphism of above mentioned groups.

Each of three elements $a,~b,~c$ in combination with unit $e$
forms a subgroup

\begin{eqnarray}
&G~~~~~\begin{array}{ccc} \supset G_+&:e,&b,\\ \supset G_-&:e,&c,\\
\supset G_0&:e,&a.\\
\end{array}&
\end{eqnarray}
Each of subgroups $G_0,~G_+,~G_-$ is the invariant subgroup
in-group $G$. The presence of the three invariant subgroups of the
second order indicates the existence of three factor-group (4:2=2)
of the second order.$$F_0:E(e,a),A(b,c),$$
$$F_+:E(e,b),A(a,c),$$ \be F_-:E(e,c),A(a,b).\ee
Here we introduced notations, common for the group theory: $E$
stands for a unit element of factor-group; $A$ stands for an
element of factor-group. Group $G$ is homomorphous to its
factor-groups $F_0$ and $F_{\pm}$. As we see from (16) the
elements of factor-group are formed as a result of unification of
certain two elements of group $G$. This kind of unification of the
elements of group $G$  indicates pair merging of energy levels
$ce_{2m},~ce_{2m+1},~se_{2m+1}$ and $se_{2m+1}$. Therefore, three
kinds of double degeneration, corresponding to the three
factor-groups  $F_0$ and $F_{\pm}$, appear in the energy spectrum.
This happens owing to the presence of parameter $l$. Therefore,
for the different values of parameter $l$ energy spectrum may be
different at least qualitatively. Let us find out how to present
combination of such a variety in energy spectrum

\begin{figure}[t]
  \centering
  \includegraphics[width=12cm]{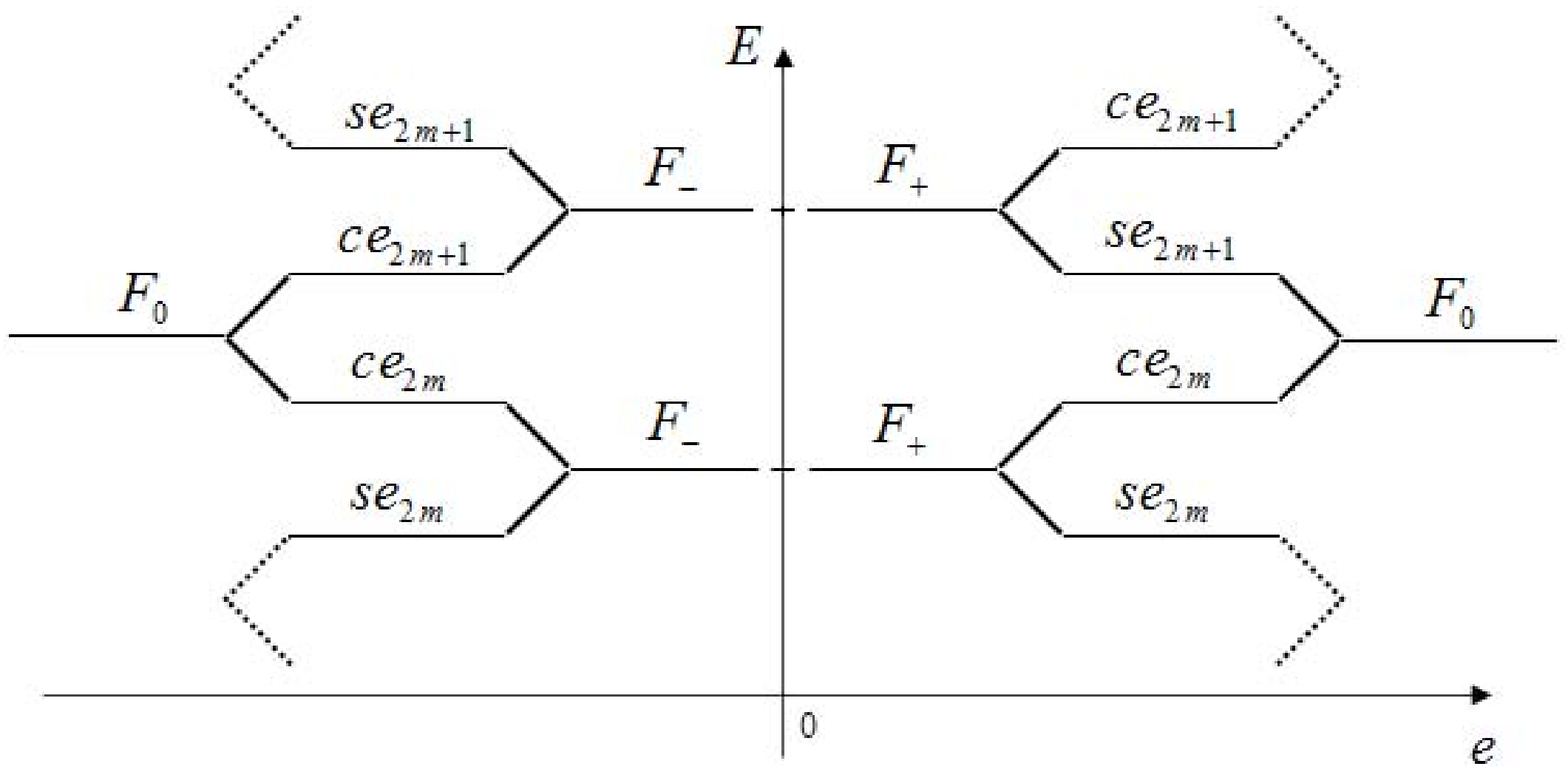}
  \caption{Fragment of energy spectrum of quantum pendulum, composed from theoretical-group consideration.}\label{fig:Fig.3}
\end{figure}

Factor-group $F_0$ is responsible for pair unity of levels of the
same symmetry relative to the center of potential well
$G(\pi-\varphi)$ (Fig.1). Quantum vibration motion appears because
of this unification. In Fig.3. these levels are arranged to the
extreme left and right from the ordinate axis.

Factor-groups $F_0$ and are responsible for pair unity of
equations $ce_n$ and $se_n(n=1,2,\ldots)$ and for formation of
clockwise and counter clockwise quantum rotation motion. In Fig.3
corresponding degenerated levels are found in both sides close to
the ordinate axis. Numerical calculations \cite{Janke-Emde-Losh}
(see Fig.4), of Mathieu characteristics $a(l)$  and $b(l)$ prove
that energy spectrum of quantum pendulum has a very complicated
form and manifests all the main features of the spectrum presented
in Fig.3.

\begin{figure}[t]
  \centering
  \includegraphics[width=12cm]{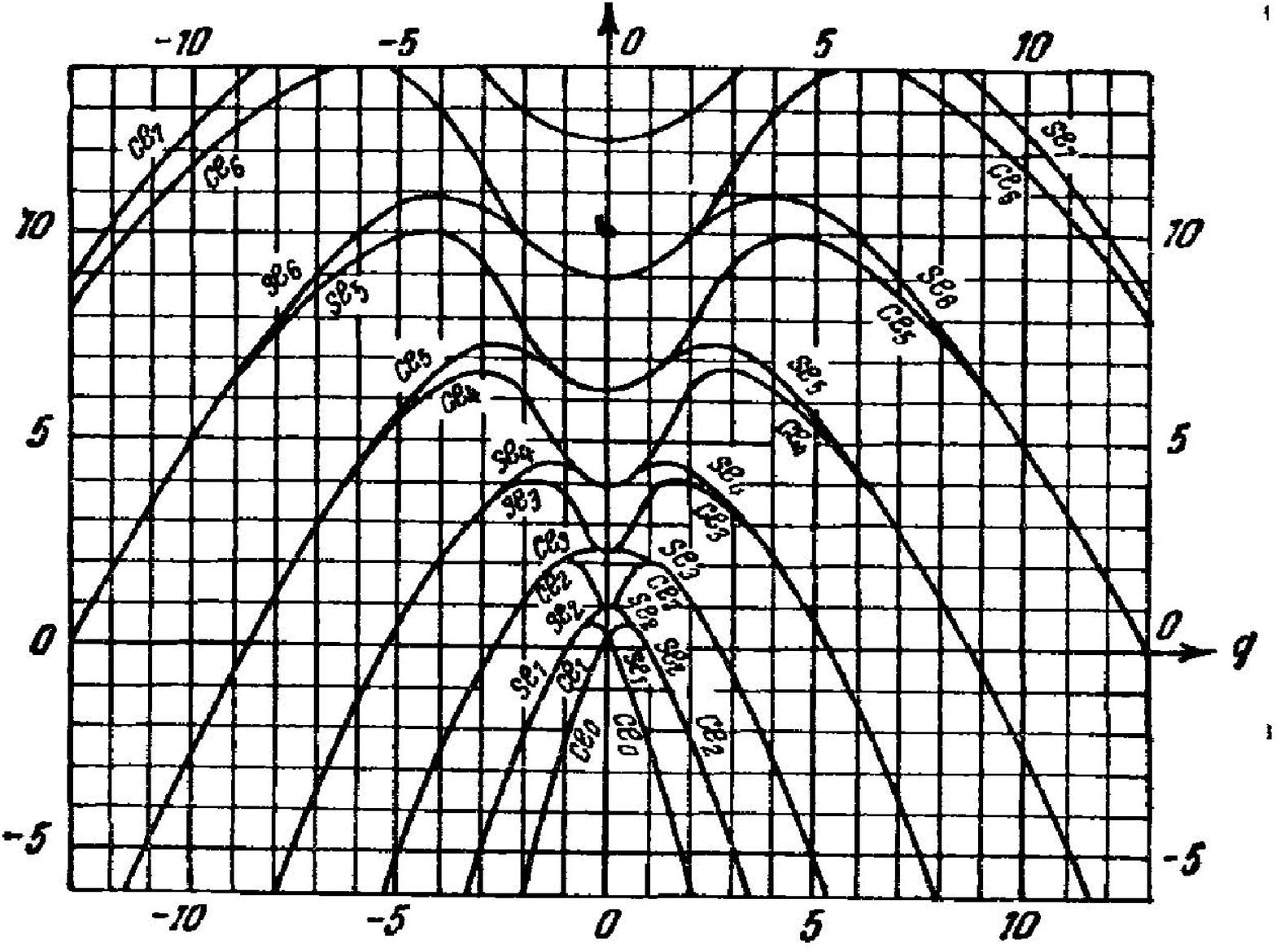}
  \caption{The group of the eigenvalues of $a_n(l)$ and $b_N(l)$ as
  functions of $l$ are plotted by numerical methods \cite{Bateman}.}\label{fig:Fig.4}
\end{figure}

Similarity of the plots presented in Fig.3 and Fig.4 is obvious.
Namely, both of them hold symmetry with respect to Y-axis. From
both sides of Y-axis same eigenstates are coupled (degenerated).
Only difference is that Fig.3 belongs to the theoretical group
analysis, while Fig.4 is plotted using numerical methods. Both of
these methods have their advantage and disadvantage. Only way to
evaluate exact positions of branch points is to use numerical
methods. From the other hand, eigenfunctions for each domain
should be defined using theoretical algebraic methods. For more
details see \cite{Chotorlishvili}.

\subsection{The Physical Problems are Reduced to Quantum Pendulum}

We have already met with one physical problem that can get reduced
to the solution of a quantum pendulum (9) - this is problem of
quantum nonlinear resonance. Now we will get to know with other
quantum-mechanical problems, that also get reduced to the solution
of quantum pendulum.

As is known \cite{Herzberg,Flygare}, one of the forms of internal
motion in polyatomic molecules is torsion oscillation which for
sufficiently large amplitudes transforms to rotational motion. In
order to describe the corresponding motion in Hamiltonian  we
assume that $\varphi$ is the angle of torsion of one part of the
molecule with respect to the other part and replace the mass $m$
by the reduced moment of inertia $I=I_1I_2/(I_1+I_2)$, where $I_1$
and $I_2$ are the inertia moments of rotation of the parts of the
molecule with respect to its symmetry axis. Thus we obtain
\cite{Gvarjaladze} \be U(\varphi)=\frac{V_0}{2}(1-\cos
n\varphi)\ee where $V_0$ defines the height of potential barrier
that separates torsion oscillations from the rotation of one part
of the molecule with respect to the other part, and n defines the
quantity of equilibrium orientations of one part of the molecule
with respect to the other part. For the molecule of ethane
$H_3C-CH_3$ , dimethylacetylene $H_3C-C\equiv C-CH_3$ and for
other organic molecules we have $n=3$ equilibrium configurations
(see Fig.5.).

\begin{figure}[t]
  \centering
  \includegraphics[width=16cm]{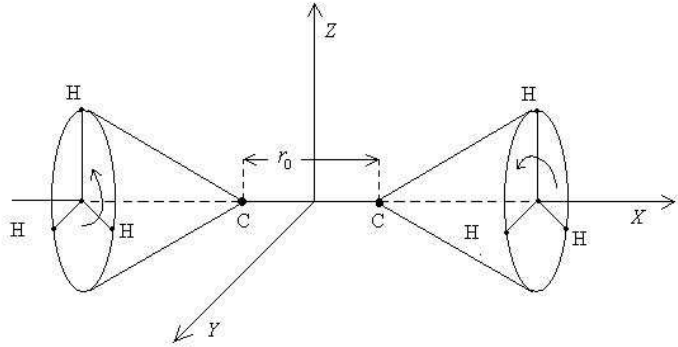}
  \caption{A schematic drawing of the molecular structure of ethane $H_3C-CH_3$.
The circular arrow shows the torsion phase  $\varphi,~r_0$ is the
equilibrium distance between two parts of
molecule.}\label{fig:Fig.5}
\end{figure}

The configuration shown in Fig.5 corresponds to an energy maximum
and is a non-equilibrium configuration (cis-configuration). Other
non-equilibrium configurations are obtained by rotating by the
angles $\frac{2\pi}{3}$ and $2\frac{2\pi}{3}.$ Equilibrium
configurations (trans-configurations) are obtained by rotating of
the angles
$\frac{\pi}{3},~\frac{\pi}{3}+\frac{2\pi}{3},~\frac{\pi}{3}+2\frac{2\pi}{3}$.

\begin{figure}[t]
  \centering
  \includegraphics[width=8cm]{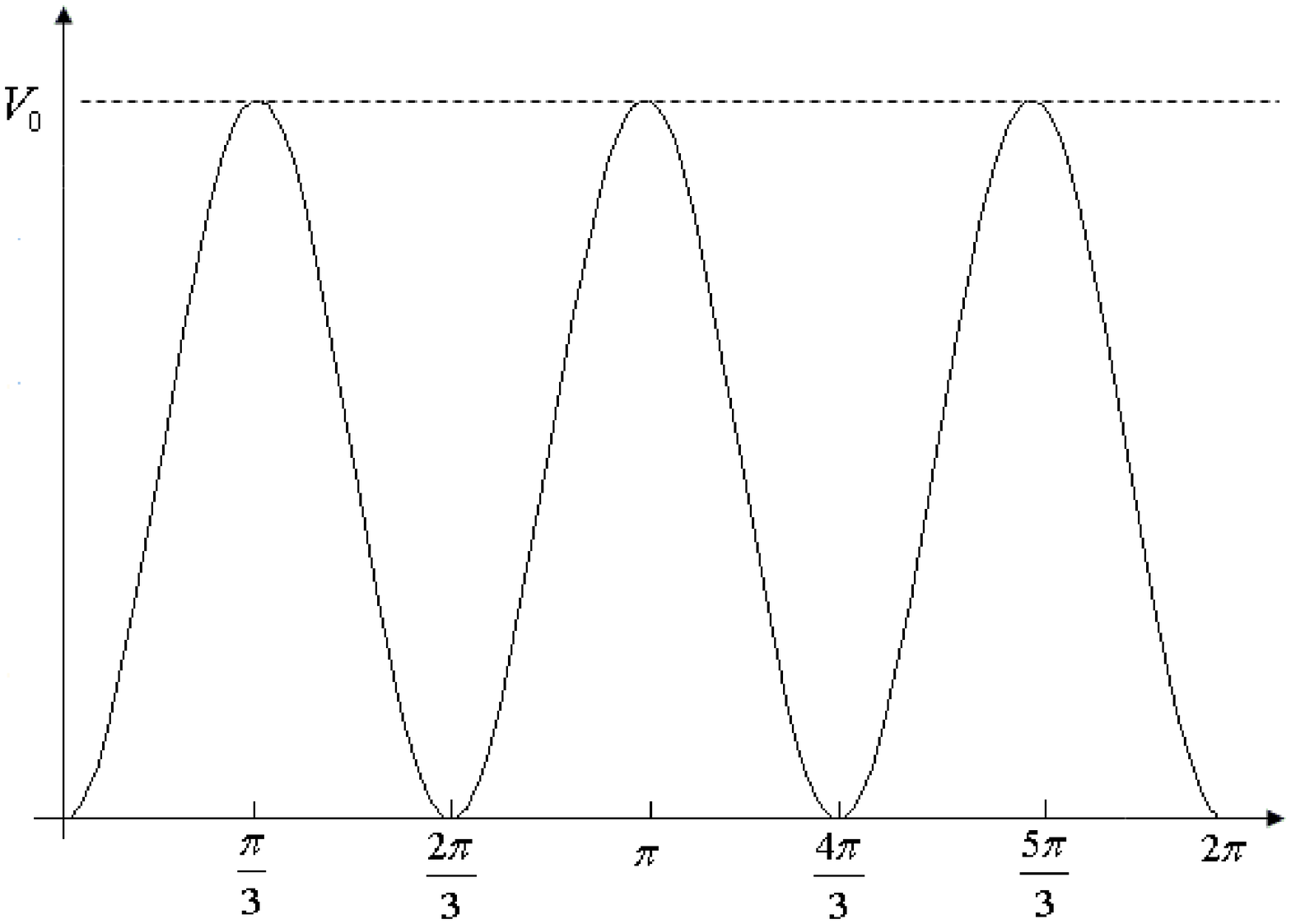}
  \caption{Potential energy curve of torsion motions in $H_3C-CH_3$.
  Phase $\varphi=0$ corresponds to equilibrium trance-configuration.}\label{fig:Fig.6}
\end{figure}

Below we give the numerical values \cite{Herzberg,Flygare} of
other parameters of some organic molecules having the property of
internal rotation. Thus for the molecule of ethane $C_2H_6$ we
have $I_1=I_2\approx 5.3\cdot 10^{-47}kg\cdot m^2$,
$V_o(C_2H_6)\approx2.1\cdot 10^{-20}J$, and for the molecule of
dimethylacetylene $C_4H_6$ we have $I_1=I_2\approx 10.6\cdot
10^{-47}kg\cdot m^2$, $V_o(C_4H_6)\approx 0.34\cdot 10^{-20}J$.

The Schrodinger equation corresponding to Hamiltonian (7) has the
form \be \frac{d^2\psi}{d\varphi ^2}+\frac{2I}{\hbar
^2}[\varepsilon_k-\frac{1}{2}V_0(1-\cos n\varphi)]\psi=0, \ee
where $\varepsilon _k$ is the eigenenergy of the $k-th$ state.
Note that $\varepsilon _k\equiv \varepsilon _k(V_o)$ is the
function of barrier height $V_o$. The condition of motion near the
separatrix (near a potential maximum) is written in the form
$\varepsilon _k\approx V_o$. If we introduce the new variable
$\alpha =\frac{n\varphi}{2}$, then equation (18) can be rewritten
as \be \frac{d^2 \psi(\alpha )}{d \alpha ^2}+[E-2l_0 \cos 2\alpha
] \psi(\alpha )=0,\ee where \be E=\frac{8I}{n^2\hbar ^2}(\epsilon
-V_0/2)\ee plays the role of energy in dimensionless units, and
the parameter \be l_0=\frac{2I}{n^2\hbar ^2}V_0 \ee is the
half-height of the barrier in dimensionless units and plays the
same role as the length of the thread does in the classical
pendulum problem.

As our next example, we consider the vibration dynamics of a
triatomic  floppy molecule: the $LiNC/LiCN$ isomerizing system
which has been extensively  studied \cite{Essers,Arranz}. This
molecule presents two stable isomers corresponding to the linear
configurations, $Li-NC$ and $Li-CN$, which are separated by a
relatively modest energy barrier. The motion in the beginning is
very floppy, and then the Li atom can easily rotate around the
$CN$ fragment

\begin{figure}[t]
  \centering
  \includegraphics[width=8cm]{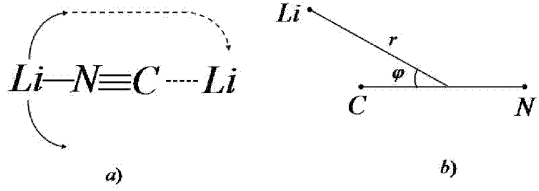}
  \caption{a) The relative motion of atom $Li$ to hard fragment $C\equiv N$. b)
Coordinates $r$ and $\varphi$ describe relative motion $Li$  to
fragment $C\equiv N$.}\label{fig:Fig.7}
\end{figure}

One can describe relative motion of  $Li$ to hard fragment
$C\equiv N$ by means of two coordinates: $\varphi$ - angle of
orientation of $Li$ with respect to hard fragments axis and $r$
distance between atom $Li$ and mass center of  fragment $C\equiv
N$. Angle $\varphi=0$ corresponds to $Li-CN$ and $\varphi=\pi$
corresponds to its isomer $CNLi$. It is easy to make sure, that
such isomerization process may be described by using the potential
(17), when $n=2$. We obtain Schrodinger equation for such case in
the same form, as we had for previous one (19).

\subsection{Degenerate states of Mathieu-Schrodinger equation.}

In the theory of Mathieu functions, the graphs of the eigenvalues
$a_n(l)$ and $b_n(l)$ as functions of $l$  are plotted by
numerical methods \cite{Bateman}. As seen from these graphs,
curves $a_n(l)$ and $b_n(l)$  merge for small  $l$, while curves
$a_n(l)$ and $b_{n+1}(l)$ merge for large $l$. It is obvious that
the merged segments of the Mathieu characteristics correspond to
the degenerate states whose existence has been mentioned above. In
this section, we will define the wave functions of degenerate
states. Below, the presence of branch points will play an
essential role in explaining the transition from the pure state to
the mixed one during the quantum investigation of the dynamics
near the classical separatrix. In what follows, we will use the
plane with coordinates  $l$, $E$. In the classical consideration,
the motion of a mathematical pendulum in a neighborhood of the
separatrix occurs when the initial kinetic energy of the pendulum
is close to the maximal potential one. It is obvious that, on the
plane $(l,E)$, to this condition there corresponds the straight
line $E=2l$. Therefore we can say that, on the plane $(l,E)$, to
non-degenerate states there corresponds a certain domain lying on
both sides of the line $E=2l$. It is in this very domain of the
change of $l$ that the system is characterized by symmetry group
$G$.
$$a) ~Degeneration ~of ~states ~at ~small ~l ~(area ~from ~the ~left ~of ~the
~separatrix ~line)$$

In the limit  $l\rightarrow 0$ the equation of the Mathieu -
Schrodinger (9) takes the form:\be
\frac{d^2\psi_n}{d\varphi^2}+E_n\psi_n=0\ee The orthonormal system
of solutions of the equation (22) consists of even and odd
solutions \be\psi_g=\cos n\varphi,~\psi_u=\sin n\varphi.\ee They
both correspond to the same energy value  $E_n=n^2$. Note that
functions (23) correspond also to the well-known asymptotic
$(l\rightarrow0)$ forms of the Mathieu functions [9]:\be
ce_n(\varphi)\rightarrow\cos
n\varphi),~se_n(\varphi)\rightarrow\sin n\varphi.\ee This means
that at the diminution of $l$ the coming together of the energy
terms with the identical $n$  takes place and for $l=0$ they are
merged together. It is necessary to find out that this confluence
happens at the point $l=0$ or at $l=l^{(n)}_-\neq0$. In this
section, below we will be concerned with finding a lower point of
the merging of terms $l^{(n)}_-$. At first let us find out what
the eigenfunctions of the degenerated states corresponding to the
level $E_n=n^2$ look like. Equation (22) is the Schrodinger
equation for free rotation in the phase plane $\varphi$. The
continuous Abelian group of two-dimensional rotations $O^+(2)$
\cite{Hamermesh} corresponds to this motion.

Since the Abelian group may have only one-dimensional irreducible
representations, the two-dimensional representation constructed in
the base of real-valued functions (23) will be reducible. Hence
functions (23) cannot be eigenfunctions of a degenerate state. To
surmount this problem we shall recollect that the eigenfunctions
for the degenerate condition can be also complex.

As is known \cite{Landau}, symmetry relative to the change of time
sign in the Schrodinger equation, accounts for the fact that the
complex-conjugate wave functions correspond to one and the same
energy eigenvalue. Therefore two complex-conjugate representations
$\psi_n(\varphi)$ and $\psi^\ast_n(\varphi)$  should be regarded
as a representation of doubled dimension. Usually, for the basis
of the irreducible representation of the group $O^+(2)$ complex
functions are assumed \cite{Hamermesh}, \be
\psi_n(\varphi)=e^{-in\varphi}.\ee Therefore, in the degenerate
domain, in view of conditions of normalization, following complex
conjugate functions should be considered as eigenfunctions \be
\psi_n(\varphi)=\frac{\sqrt{2}}{2}e^{-in\varphi},~\psi^\ast_n(\varphi)=
\frac{\sqrt{2}}{2}e^{in\varphi}.\ee Let us remark that group
$O^+(2)$ is isomorphic to subgroup $G_-$ (24). The element of the
symmetry $c=G(\varphi\rightarrow\pi+\varphi)$ of the subgroup
$G_-$ provides recurrence of the phase variation after each period
and consequently the symmetry $G_-$ characterizes the condition of
motion similar to the classical rotary motion. However, to use
only the argument of symmetry is not sufficient for finding the
coordinates of the branching point $l^{(n)}_-$. Below to find
these points we use the secular perturbation theory. So, at $l=0$
we have doubly degenerate states with the  wave functions (26).
Let us find out, whether the perturbation \be V(l,\varphi)=l\cos
2\varphi,~~l\ll1\ee can remove the existing degeneration. It is
known that, first order terms of the perturbation theory for the
energy eigenvalues and the exact functions of zero approximation
for double degenerate levels look like \cite{Landau}
$$E^{(1)}_{0\pm}=\frac{1}{2}[(V_{11}+V_{22})\pm\sqrt{(V_{11}+V_{22})^2+4\cdot|V_{12}|^2}],$$
$$\psi_n^{\pm}=\psi^{(0)}=C_1\cdot\psi_1^0+C_2\cdot\psi_2^0,$$
$$C_1^{(0)}=\bigg\{\frac{V_{12}}{2|V_{12}|}\bigg[1\pm\frac{V_{11}-V_{22}}{\sqrt{(V_{11}+V_{22})^2+4\cdot|V_{12}|^2}}\bigg]\bigg\}^{1/2}$$
\be
C_2^{(0)}=\pm\bigg\{\frac{V_{21}}{2|V_{12}|}\bigg[1\mp\frac{V_{11}-V_{22}}{\sqrt{(V_{11}+V_{22})^2+4\cdot|V_{12}|^2}}\bigg]\bigg\}^{1/2}\ee
where the index in brackets corresponds to the order of the
perturbation theory, matrix elements of the perturbation (27)
$V_{ik}(i,k=1,2)$ are calculated by using of functions (26) of the
degenerate state of the non-perturbed Hamiltonian. Taking into
account expressions (26) we shall calculate the matrix elements:
$$V_{11}=l\int_0^\pi \psi_n^\ast(\varphi)\cdot\psi_n(\varphi)\cos 2\varphi d\varphi=0,~V_{22}=0,$$
\begin{displaymath}
V_{12}=l\int\limits_{0}^{\pi}\Psi_{0}^{2}(\varphi)\cdot\cos2\varphi
d\varphi=\left\{\begin{array}{ll} 0 & \textrm{if $n\neq1$}\\
\frac{l\pi}{4} &  \textrm{if $n=1$}
\end{array} \right.
\end{displaymath}
After substitution of those matrix elements in the expressions
(28) for the eigenvalues and exact eigenfunctions we shall obtain:
\be E_\pm^{(1)}=\pm\frac{l\pi}{4},~~\psi_{n=1}^+=\cos
\varphi,~~\psi_{n=1}^-=-i\sin \varphi.\ee Thus, the exact wave
functions (29) of the non-degenerate state only for $n=1$ coincide
with the Mathieu function in the limit $l\rightarrow 0$ (24).

The perturbation  $V(l,\varphi)$ removes degeneration only for the
state n=1. Therefore it is only for the state $n=1$ that the
spectrum branching occurs at the point $l=0$, which agrees with
numerical calculations given in the form of diagrams (see Fig.
4.). It can be assumed that in the case of diminishing $l$, the
merging of energy terms for states $n\neq 1$ takes place at the
point at which the states are still defined by the Mathieu
functions and not by their limiting values (24). Wave functions
for degenerate states $l\neq 0,~n\neq 1$ can be composed from the
Mathieu functions by using the same arguments as have been used
above in composing the wave functions for $l\rightarrow0)$. As a
result, we obtain
$$\psi_+^{2m+1}=\psi_n(l,\varphi)=\frac{\sqrt{2}}{2}(ce_n\varphi\pm ise_n\varphi),~n=2m+1,$$
\be
\psi_-^{2m+2}=\psi_n^\ast(l,\varphi)=\frac{\sqrt{2}}{2}(ce_n\varphi\pm
ise_n\varphi),~n\neq1,~l\neq0,~n=2m+2.\ee Let us assume that at
$l_n=l^{(n)}_-$ the removal of degeneration for the $n-th$ energy
term happens.

\begin{figure}[t]
  \centering
  \includegraphics[width=16cm]{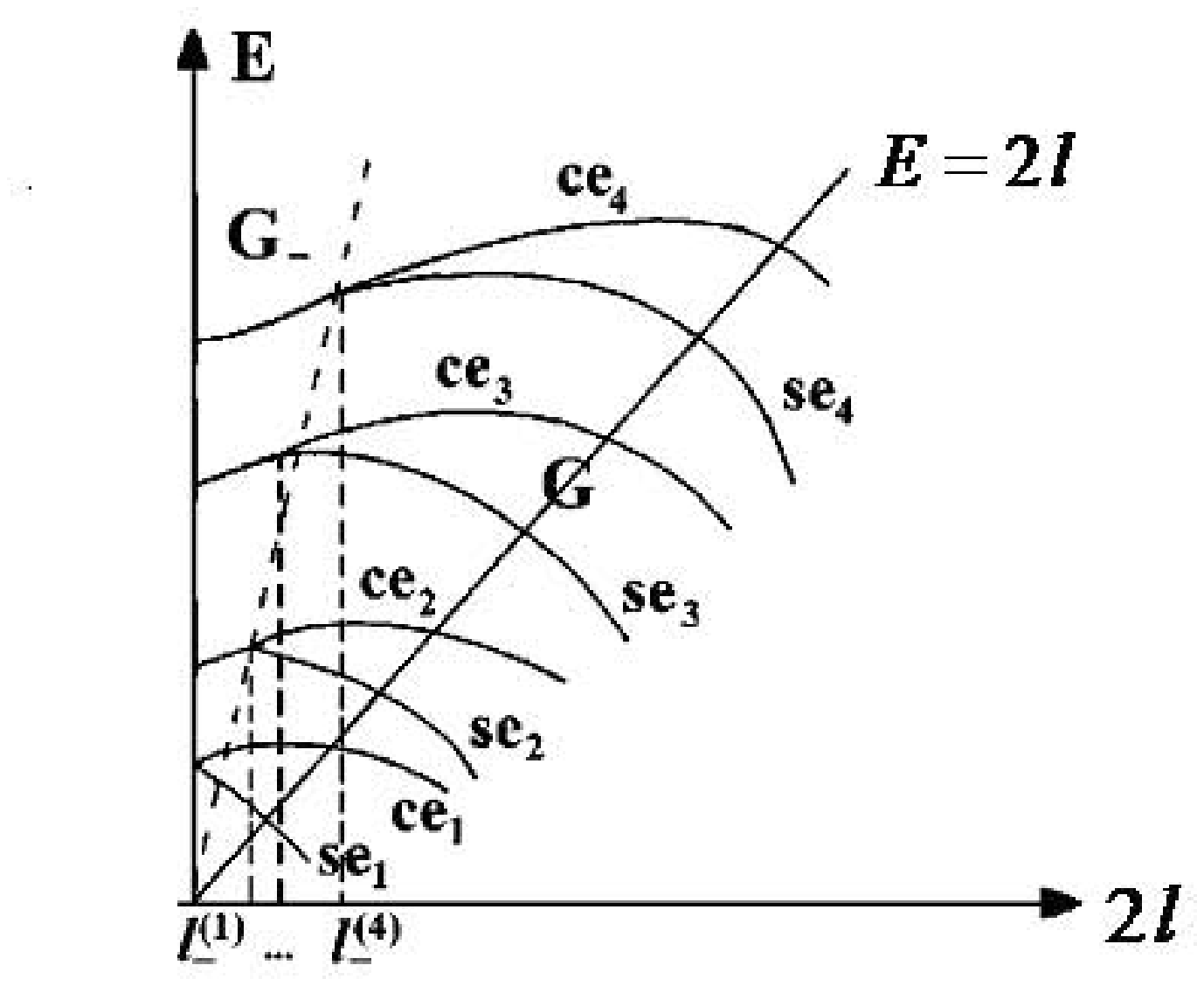}
  \caption{The energy levels as function of parameter $l$ on the plane
  $(E, l)$ on the left area from  the separatrix line. The points
  of the branching of curves represent the boundaries between
  degenerate and non-degenerate states.}\label{fig:Fig.8}
\end{figure}
$$b) ~Degenerate ~state ~at ~major ~l. ~The ~area ~on ~the ~right ~of ~the
~separatrix ~line.$$

With the increasing of $l$ the particle can be trapped in a deep
potential well ($V=l\cos 2\varphi,~0<\varphi<\pi$ Fig. 1.), and
perform oscillatory motion. Properties of wave functions of
quantum oscillator near to the bottom of the well are well known.
This is the alternation of even and odd wave functions relative to
the center of the potential well $\pi/2$ and presence of zeros in
wave functions. With the help of the third column of Table 1 it is
possible to write symmetry conditions close to $\pi/2$:
$$ce_m(\frac{\pi}{2}+\varphi)=(-1)^mce_m(\frac{\pi}{2}-\varphi)$$
\be se_m(\frac{\pi}{2}+\varphi)=(-1)^{m+1}
se_m(\frac{\pi}{2}-\varphi)\ee i.e.
$ce_{2m}(\varphi),~se_{2m+1}(\varphi)$ are even functions and
$se_{2m}(\varphi),~ce_{2m+1}(\varphi)$ are odd functions.
Functions
$ce_{2m}(\varphi),~se_{2m+1}(\varphi),~ce_{2m+1}(\varphi)$ and
$se_{2m+1}(\varphi)$  have $m$ real zeros between $\varphi=0$ and
$\varphi=\pi/2$ (not considering zeros on edges). The existing
alternation of states (Fig. 8.) in area along the line of the
separatrix is conditioned by the properties of states at the small
$l$. With the help of the expressions (31) it is possible to
determine easily, that in the spectrum of the states along the
line $E=2l$ two (instead of one) even states alternate with odd
states and so on. To get the alternation, caused now by properties
at major $l$, two even conditions must degenerate in one even and
two odd - in one odd. So we come to the conclusion, that two
levels with wave functions $ce_{2m}(\varphi)$ and
$se_{2m+1}(\varphi)$ coming nearer amalgamate in one level, and
the following two levels $ce_{2m}(\varphi)$ and
$se_{2m+1}(\varphi)$  also in one level. The levels obtained in
this way will be doubly degenerated. It can be assumed that with
the growth of $l$ the states defined by the symmetry group $G$
transform to the states with the symmetry of an invariant subgroup
$G_+$ (15). This transformation takes place at the merging point
of non-degenerate terms $l_n=l^{(n)}_+$. Recall that subgroup
$G_+$ contains two elements: the unit element e and the reflection
element with respect to the symmetry center of the well
$b=G(\varphi\rightarrow\pi-\varphi)$. Complex wave functions of
the area of degenerate states, with the symmetry of the invariant
subgroup $G_+$, can be composed of pairs of functions of merged
states in the same manner as we have done above for the area of
small $l$ for states with the symmetry of $G_-$.

Not iterating these reasons, we shall write complex wave functions
corresponding to the degenerated states in the form
$$\xi^{\pm}_{2n}(\varphi)=ce_{2n}(\varphi)\pm se_{2n+1}(\varphi)~~even~state$$,
\be \xi^{\pm}_{2n+1}(\varphi)=ce_{2n+1}(\varphi)\pm
se_{2n+2}(\varphi)~~odd~state\ee

In the base of complex wave functions $\xi^{\pm}_{2n}$ and
$\zeta^{\pm}_{2n+1}$ the indecomposable representation of the
subgroup $G_+$ (15), is realized. Evenness of the wave functions
$\xi$ and $\zeta$ with respect to the transformation
$b=G(\varphi\rightarrow\pi-\varphi)$ of the subgroups $G_+$
characterizes an important property of wave functions evenness of
the quantum oscillatory process. The results, obtained in this
section, are plotted in Fig. 9.

\begin{figure}[t]
  \centering
  \includegraphics[width=16cm]{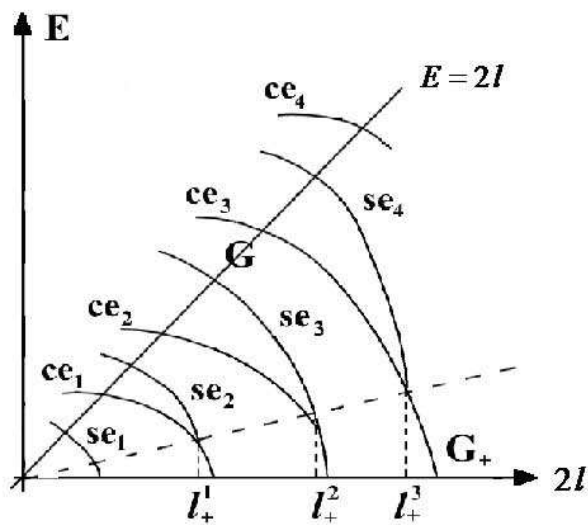}
  \caption{ Energy levels as a function of the parameter $l$ on
  the plane $(E,l)$ to the area right from the separatrix
  line. The points of branching of curves represent
  degeneration points of terms in this area. }\label{fig:Fig.9}
\end{figure}

Figures 8 and 9 supplement each other: in the field of
intersection with the separatrix the curves of the Figs. 8 and 9
are smoothly joined. So, we shall add up outcomes obtained in this
section. The Mathieu-Schrodinger equation has an appointed
symmetry. The transformations of the symmetry of the Mathieu
functions form group $G$, which is isomorphic to the quaternary
group of Klein. To this symmetry on a plane $(E,l)$ corresponds
the appointed area along the line of the separatrix $E=2l$,
containing non-degenerated energy terms. This area is restricted
double sided by the areas of degenerate states, which are
characterized by the symmetry properties of the invariant
subgroups $G_-$ and $G_+$, respectively. The boundaries of these
areas are defined by the branching points of energy terms existing
both on the right and on the left of the separatrix.

The area of degenerate states is the quantum-mechanical analogs of
two forms of motion of the classical mathematical pendulum-rotary
and oscillatory. Comparing results of quantum reviews with
classical, we remark that these two conditions of motion at
quantum reviewing are divided by the area of a finite measure,
whereas at the classical reviewing measure the separatrix is equal
to zero.

\subsection{Quantum analog of the stochastic layer.}

In the case of Hamiltonian systems, performing a finite motion, a
stochastic layer formed in a neighborhood of the separatrix under
the action of an arbitrary periodic perturbation is a minimal
phase space cell that contains the features of stochasticity
\cite{Sagdeev}. In this section we shall try to find out what can
be considered as the quantum analog of the stochastic layer.

Let us assume that the pumping amplitude is modulated by the slow
variable electromagnetic field. The influence of modulation is
possible to take into account by means of such replacement in the
Mathieu-Schrodinger equation (9),\be l\rightarrow l_0+\Delta l\cos
\nu t,~~\Delta l<l_0.\ee Here $\Delta l$ stands for the amplitude
of modulation in dimensionless unit (see (10)), $\nu$ is the
frequency of modulation. We suppose, that the slow variation of l
can embrace some quantity of the branching points on the left and
on the right of separatrix line (Figs. 8,9)\be \Delta
l\geq|l_+^n-l_-^n|,~~n=1,2,...N.\ee As a result of replacement
(33) in the Hamiltonian (7), we get \be
\hat{H}=\hat{H}_0+\hat{H}'(t),\ee \be \hat{H}'(t)=\Delta l\cos
2\varphi\cos \nu t, \ee where $H_0$ is the universal Hamiltonian
(7) and $\hat{H}'(t)$ is the perturbation appearing as a
consequence  of pumping modulation.

It is easy to see, that the matrix elements of perturbation (36)
$\hat{H}'(t)$ for non-degenerate states equal zero. Really, having
applied expansion formulas of the Mathieu functions in the Fourier
series \cite{Janke-Emde-Losh} it is possible to show
\be<ce_n|\hat{H}'(t)|se_n>\sim\int_0^{2\pi}ce_n(\varphi)\cos
2\varphi se_{2n}(\varphi)d\varphi=0 \ee simultaneously for the
even and odd n. The expressions of the selection rules (37) will
be fulfilled for values $l$ from the area $l^n_-\leq l\leq ^n_+l$
Transitions between levels cannot be conditioned by time-dependent
perturbation (36). It is expedient to include perturbation in the
unperturbed part of the Hamiltonian. The Hamiltonian, obtained in
such way, is slowly depending on the parameter $l$. So, instead of
(35) and (36) for the non-degenerated area $G$ we get the
Hamiltonian in the form \be
\hat{H}=-\frac{\partial^2}{\partial\varphi^2}+l(t)\cos
2\varphi,\ee \be l(t)=l_0+\Delta l\cos \nu t.\ee There arises the
situation in which the system slowly "creeps" along the Mathieu
characteristics and, in doing so, encloses the branching points on
the left $l^n_-$ or on the right $l^n_+$. a) \textit{~Irreversible
~"creeping" ~of ~energy ~term ~populations ~due ~to ~the
~influence ~of ~a ~measuring ~arrangement.}

According to the general rules of quantum mechanics, probabilities
that the system will pass to the eigenstate of another area are
defined by the coefficients of expansion of the wave function of
one area into the eigenfunctions of another area. Let us assume
that, initially, the system was in one of the eigenstates from the
non-degenerate area $G$, for example, in the state $ce_{2n}$.
After a quarter of the modulation period $T/4$ (where $T=2p/n$),
having passed through the point $l^n_-$, the system finds itself
in the degenerated area $G_-$. In this case the system will pass
to degenerate states $\psi^{\pm}_{2n}$ with probabilities,
$$P(ce_{2n}\rightarrow \psi_{2n}^\pm)=|\frac{1}{\pi}\int_0^{2\pi}ce_{2n}(\varphi)\psi_{2n}^{\pm\ast}(\varphi)d\varphi|^2=$$
\be
=\frac{1}{2\pi}|\int_0^{2\pi}ce_{2n}(\varphi)(ce_{2n}(\varphi)\pm
ise_{2n}(\varphi))^\ast d\varphi|^2=1/2.\ee For deriving (40) we
used the condition of normalization (12) and orthogonality [9] \be
\int_0^{2\pi}ce_k(\varphi) se_{l+1}(\varphi)
d\varphi=0,~l,k=0,1,2,...).\ee

The passage (40) is based on the assumption of having a deep
physical sense. As is generally known, in quantum mechanics
symmetry with respect to both directions of time is expressed in
the invariance of the Schrodinger equation with respect to the
variation of the sign of time $t$ and simultaneous replacement
$\psi$ by way of $\psi^\ast$. However, it is necessary to remember
that this symmetry concerns only the equations, but not the
concept of a measurement playing a fundamental role in the quantum
mechanics \cite{Landau,Kaempffer}. "Measurement" is understood as
the process of interaction of the quantum system with the
classical object usually called "instrument." Under the measuring
arrangement, consisting of the analyzer and detector, one must not
imagine the laboratory's instrument. In our case, the role of the
instrument plays in our case the modulating field, which is
capable to "drag" the system through the branching points. When
passing through the branching point from one area to another, the
state remains unchanged. However, being an eigenstate in one area,
it will not be an eigenstate in another. At the passage through
branching points there occurs a spectral expansion of the initial
wave function belonging to the region of one symmetry over the
eigenfunctions belonging to the region of another symmetry. The
presence only of the analyzer reserves a pure state and the
process remains reversible. So, passage through the branching
point plays role of analyzer. Further we shall assume the presence
of the detector, defining which of the states $\psi_n^+$ or
$\psi_n^-$ is involved in passage. The transition of the system to
various states defined by probabilities (40) is fixed by means of
the action of the detector. The presence of the detector is
expressed formally in averaging with respect to phase and
neglecting the interference term usually appearing in the
expression for a distribution function. As a result of averaging
the partial loss of information about the condition of the system
takes place and a mixed state is generated. In our problem role of
detector will play effect self-chaotization, which appeared in
degenerate region (see next subsection).

As is follows from (40), after the quarter period degenerated
rotary states $\psi_{2n}^+$ and $\psi_{2n}^-$ will be occupied
with the identical probability. After the half period
$\frac{1}{2}T$ the system again appears in the area $G$ going
through the branching point $l_-^n$ in the reverse direction. At
the same time there appear probabilities, of the transition into
the states $ce_n,~se_n$, and both of them are distinct from zero
\be P(\psi_{2n}^\pm \rightarrow
ce_{2n})=\frac{1}{2}|\frac{1}{\pi}\int_0^{2\pi}(ce_{2n}(\varphi)\pm
ise_{2n}(\varphi))ce_{2n}(\varphi)d\varphi|^2=\frac{1}{2},\ee \be
P(\psi_{2n}^\pm \rightarrow
se_{2n})=\frac{1}{2}|\frac{1}{\pi}\int_0^{2\pi}(ce_{2n}(\varphi)\pm
ise_{2n}(\varphi))se_{2n}(\varphi)d\varphi|^2=\frac{1}{2}.\ee Here
we have used again normalization (12) and orthogonality relations
(41). It is easy to write transition probability from $ce_{2n}$
into one of the degenerated states $\psi_{2n}^\pm$ and back in the
$ce_{2n}$
$$P(ce_{2n}\leftrightarrow ce_{2n})\equiv P(ce_{2n}\rightarrow\psi_{2n}^\pm\rightarrow ce_{2n})=$$
\be = P(ce_{2n}\rightarrow \psi_{2n}^+)P(\psi_{2n}^+\rightarrow
ce_{2n})+P(ce_{2n}\rightarrow \psi_{2n}^-)P(\psi_{2n}^-\rightarrow
ce_{2n}).\ee Here the first summand corresponds to the passage
through the degenerated state $\psi_{2n}^+$ and the second one to
the passage through $\psi_{2n}^-$. It is easy to see with the help
of previous computations (40), (42), and (43) that contributions
of these passages are identical and individually equal to $1/4$.
Therefore finally we have\be P_-(ce_{2n}\leftrightarrow
ce_{2n})=\frac{1}{2}.\ee Similarly it may be shown that transition
probability from the state $ce_{2n}$  in one of the degenerated
states $\psi_{2n}^\pm$ and back in the area $G$, in the state
$se_{2n}$  by means of going through the point $l_-^n$ is $$
P(ce_{2n}\leftrightarrow se_{2n})= P(ce_{2n}\rightarrow
\psi_{2n}^+)P(\psi_{2n}^+\rightarrow se_{2n})+$$ \be
+P(ce_{2n}\rightarrow \psi_{2n}^-)P(\psi_{2n}^-\rightarrow
se_{2n})=\frac{1}{2}\cdot\frac{1}{2}+\frac{1}{2}\cdot\frac{1}{2}=1/2.\ee
Thus, the system being at the initial moment in the eigenstate
$ce_{2n}$, at the end of half-period of modulation appears in the
mixed state $\rho_{2n}$ in which the states $ce_{2n}$ and
$se_{2n}$ are intermixed with identical weight, and corresponding
levels are populated with identical probabilities. After the
expiration of quarter of cycle the system will pass from the area
$G$ (the state $\rho_{2n}$) in the area  $G_+$, going through the
point $l_n^+$. In passages from the area $G_+$ four states take
part $\xi_{2n}^\pm = \frac{1}{\sqrt{2}}(ce_{2n}\pm ise_{2n+1})$
and $\zeta_{2n-1}^\pm = \frac{1}{\sqrt{2}}(ce_{2n-1}\pm
ise_{2n})$. So, with taking into consideration the above mentioned
for the probabilities of transitions we get \be
P(\rho_{2n}\rightarrow
\xi_{2n}^\pm)=\frac{1}{4\pi}|\int_0^{2\pi}(ce_{2n}(\varphi)+se_{2n}(\varphi))(ce_{2n}(\varphi)\mp
ise_{2n+1}(\varphi))d\varphi|^2=\frac{1}{4},\ee \be P(\rho_{2n}
\rightarrow
\zeta_{2n}^\pm)=\frac{1}{4\pi}|\int_0^{2\pi}(ce_{2n}(\varphi)+se_{2n}(\varphi))(ce_{2n-1}(\varphi)\mp
ise_{2n}(\varphi))d\varphi|^2=\frac{1}{4},\ee

For deriving the last expressions in addition to the normalization
conditions we have used the orthogonality conditions \be
\int_0^{2\pi}ce_n(\varphi)ce_m(\varphi)d\varphi=\int_0^{2\pi}se_{n+1}(\varphi)se_{m+1}(\varphi)d\varphi=0,~m\neq
n.\ee On the basis of (47) and (48) we conclude, that after the
time $\frac{3}{4}T$, system will be in the area $G_+$ in one of
four oscillatory states $\xi_{2n-1}^\pm$ and $\zeta_{2n-1}^\pm$
with the identical probability equal to $1/4$.

After one cycle $T$ the system gets back in the area $G$, from
which it started transition from the level $ce_{2n}$. Upon
returning, four levels $ce_{2n},~se_{2n},~ce_{2n-1}$ and
$se_{2n+1}$ will be involved. Calculating probabilities of
passages from the oscillatory state of the area $G_+$, to these
four levels we shall obtain \be P(\xi_{2n}^\pm\rightarrow
ce_{2n})=P(\xi_{2n}^\pm\rightarrow se_{2n+1}=1/2,\ee \be
P(\xi_{2n-1}^\pm\rightarrow se_{2n})=P(\zeta_{2n-1}^\pm\rightarrow
se_{2n-1}=1/2.\ee The probability of passages from the
nondegenerated area to the area $G_+$ in one of the oscillatory
states $\xi_{2n}^\pm,~\zeta_{2n-1}^\pm$ and back in the area $G$
will be:
$$P_+(\rho_{2n}\leftrightarrow se_{2n})=P(\rho_{2n}\rightarrow \xi_{2n}^\pm)P(\xi_{2n}^+\rightarrow se_{2n+1})+$$
\be +P(\rho_{2n}\rightarrow \xi_{2n}^-)P(\xi_{2n}^-\rightarrow
se_{2n+1})=\frac{1}{4}\cdot\frac{1}{2}+\frac{1}{4}\cdot\frac{1}{2}=\frac{1}{4}.\ee
Similarly it is possible to show \be P(\rho_{2n}\leftrightarrow
ce_{2n})=P(\rho_{2n}\rightarrow se_{2n})=P(\rho_{2n} \rightarrow
ce_{2n-1})=1/4.\ee Thus, after the lapse of time $T$ four levels
of the nondegenerate area $G$ will be occupied with the identical
probabilities $1/4$ (Fig. 10.) The motion of the system upwards on
energy terms will cease upon reaching the level for which the
points of the branching in Fig. 10 are on the distance at which
the condition (34) no longer is valid. The motion of the system
downwards will be stopped upon reaching the zero level. If the
system at the initial moment is in the state $2n=N/2$, then after
$N/2$ cycles of modulation all $N$ levels will be occupied. It is
easy to calculate a level population for the extremely upper and
extremely lower levels. Really, the level population for extreme
levels is possible to define with the help of a Markov chain
containing only one possible trajectory in the spectrum of Mathieu
characteristics:
$$P(ce_{N/2},t_0;ce_{N/2+1},t_0+\frac{T}{2},\dots ce_N,t_0+N\frac{T}{2})=$$
$$P(se_N,t_0+N\frac{T}{2}\leftarrow ce_{N-1},t_0+(N-1)\frac{T}{2})...$$
\be P(ce_{N/2+1},t_0+T\leftarrow
se_{N/2+1},t_0+\frac{T}{2})P((se_{N/2+1},t_0+\frac{T}{2}\leftarrow
ce_{N/2+1},t_0),\ee where $t_0$ is an initial time. Here, when
discussing the transition probabilities from one state to another,
we also use a time argument. It is possible to write a similar
chain of level population for the extremely lower level. As the
probabilities of passages, included in the right side of (54) by
way of factors, are equal to $1/2$, then probabilities of an
extreme level population will be $(1/2)^{N/2}$. As to the
Markovian chain for non-extreme levels, it has a cumbersome form
and we do not give it here.

\begin{figure}[t]
  \centering
  \includegraphics[width=16cm]{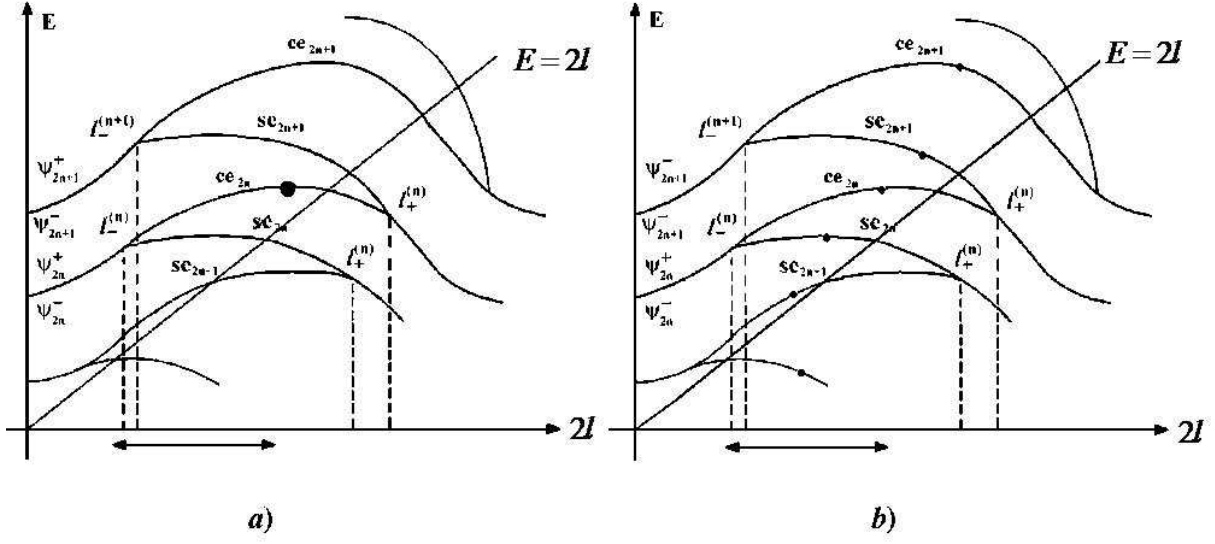}
  \caption{The fragment of the energy terms, participating in passages
calculated in the text. a) The initial state. The particle is in
the state of $ce_{2n}$. b) The final state. Levels which are
affected by a change of the field amplitude get
populated.}\label{fig:Fig.10}
\end{figure}
$$b) ~Selfchaotization ~produced ~by ~a ~big ~phase ~"incursion" ~of ~the
~probability ~amplitude.$$

Different from the area of non-degenerate states $G$, in the areas
of degenerate states $G_-$ and $G_+$, the non-diagonal matrix
elements of perturbation $\hat{H}'(t)$ (36) are not zero. \be
H^\prime_{+-}=H^\prime_{-+}=<\psi^+_{2n+1}|\hat{H^\prime(t)}|\psi^-_{2n+1}>
\sim\Delta l
\int\limits_{o}^{2\pi}\psi^+_{2n+1}\psi^{-\ast}_{2n+1}\cos
2\varphi d\varphi \not= 0. \ee where the wave functions $\psi_\pm$
have been defined previously by (30). Here, for the brevity of the
notation, we omit the upper indices indicating the quantum state.
An explicit dependence of $\hat{H}'(t)$ on time given by the
multiplier $\cos \nu t$ is assumed to be slower as compared with
the period of passages from one degenerate state to another
produced by the non-diagonal matrix elements $H'_{+-}$. Therefore
below, perturbations $\hat{H}'(t)$ will be treated as
time-independent perturbations able to produce the above-mentioned
passages. Therefore in the area of degenerate states the system
can be found in the time-dependent superposition state
\cite{Landau,Bohm}:
 \be \psi_{2n}(t)=C^+_n(t)\psi^+_{2n}+C^-_n(t)\psi^-_{2n}.\ee
Probability amplitudes $C^\pm_n(t)$ are found by means of the
following fundamental quantum-mechanical equation expressing the
causality principle:
$$-i\hbar\frac{dC^+_n}{dt}=(E_{on}+H^\prime
_{++})C^+_n+H^\prime_{+-}C^-_n,$$ \be
-i\hbar\frac{dC^-_n}{dt}=H^\prime _{+-}C^+_n+(E_{on}+H^\prime
_{--})C^-_n.\ee

It is easy show that in the case of our problem it should be
assumed that $H'_{++}=H'_{--}$ and $H'_{+-}=H'_{-+}$. Let us
investigate changes that occurred in the state of the system
during time $\Delta T$ while the system was in the area $G_-$,
(i.e., during the time of movement to the left from $l_-^n$ and,
reversal, to the right to $l_-^n$). It will be assumed that $T$ is
part of the period of modulation $T$. For arbitrary initial values
the system of equations (57) has a solution
$$C^+_n(t)=\frac{C_+(0)+C_-(0)}{2}exp[\frac{-i}{\hbar}(E-H')t]+\frac{C_+(0)-C_-(0)}{2}exp[\frac{i}{\hbar}(E-H')t],$$
\be
C^-_n(t)=\frac{C_+(0)+C_-(0)}{2}exp[\frac{-i}{\hbar}(E-H')t]-\frac{C_+(0)-C_-(0)}{2}exp[\frac{i}{\hbar}(E-H')t],\ee
where $E\rightarrow E_0+H'_{++},~E\rightarrow
E_0+H'_{--},~H'_{+-}\rightarrow H'.$

After complementing $\hat{H}'(t)$ (58) with the factor $\cos \nu
t$, we can take into consideration also a slow time dependent
change of perturbation $(H'\rightarrow H'\cos \nu t).$

Let motion begin from the state $\psi^-_{2n}$ of the degenerate
area. Then as the initial conditions we take \be
C^-_n(0)=1,~C^+_n(0)=0\ee as initial conditions. Substituting (59)
into (58), for the amplitudes $C_\pm(t)$ we obtain
$$C_-(t)=e^{(i/\hbar)Et}\cos \omega t,$$ \be C_+(t)=-ie^{(i/\hbar)Et}\sin \omega t,~\omega=\frac{H'}{\hbar}.\ee
Now, using (60), for the distribution $|\psi(t)|^2$ (56) we obtain
$$|\psi(t)|^2=\cos^2(\frac{H'}{\hbar}t)|\psi_-(t)|^2+\sin^2(\frac{H'}{\hbar}t)|\psi_+(t)|^2-$$
\be
-\frac{1}{2}\sin(2\frac{H'}{\hbar}t)[\psi_+(t)\psi^\ast_-(t)-\psi_-(t)\psi^\ast_+(t)].\ee
In the expression for $|\psi(t)|^2$ the first two terms correspond
to the transition probabilities $-\rightarrow +$ and $-\rightarrow
+$, respectively, while the third term corresponds to the
interference of these states. Distribution (61) corresponds to a
pure state.

Note that (like any other parameter of the problem) the value $H'$
contains a certain small error $\delta H'\ll H'$, which during the
time of one passage $2\pi\hbar/H'$ leads to an insignificant
correction of the phase $2\pi(\delta H'/H')$. However, during the
time $T$ a phase incursion takes place and a small error $\delta
H'$ may lead to uncertainty of phase $\sim(\delta H'/\hbar)\Delta
T$, which may turn out to be of order $2\pi$. In that case the
phase becomes random. Therefore by the moment $\Delta T$ the
distribution takes the form that can be obtained from (61) by
means of averaging with respect to the random phase
$\alpha=(\delta H'/\hbar)\Delta T$.

Hence, after averaging expression (61), equating the interference
term to zero and taking into account that
$$\overline{\sin^2[(H'/\hbar)t]}=\overline{\cos^2[(H'/\hbar)t]}=1/2$$
we get
\be
|\psi|^2=\bar{|\psi(t)|^2}=\frac{1}{2}(|\psi_+|^2+|\psi_-|^2),
\ee
where the stroke above denotes the averaging with respect to time.
The obtained formula (62) is the distribution of a mixed state,
which contains probabilities of degenerate states $|\psi_\pm|^2$
with the same weights $1/2$. The assumption that a large phase is
a random value that, after averaging, makes the interference term
equal to zero is frequently used in analogous situations
\cite{Bohm}.

Thus we conclude that if the system remains in the areas $ G_\pm $
of degenerate states for a long time, $ \Delta T\gg 2\pi\hbar /
H',~ \Delta T\approx 2\pi \hbar/\delta H' $ during which the
system manages to perform a great number of passages, then in the
case of a passage to the nondegenerate area $G$ the choice of
continuation of the path becomes ambiguous. In other words, having
reached the branch point, the system may with the same probability
continue the path along two possible branches of the Mathieu
characteristics. The error $H'$ is evidently connected with the
error of the modulation amplitude value. It obviously follows
that, when passing the branch point, the mixed state (62) will
transform with a $1/2$ probability to the states $ce$ and $se$, as
shown in formulas (42) and (43). Analogously, we can prove the
validity of all subsequent formulas for the passage probabilities
(44),(45).
$$c) ~Kinetic ~stage ~of ~evolution$$

We have already met with one physical problem that can get reduced
to the solution of a quantum pendulum (9) - this is a problem of
quantum nonlinear resonance. Now we will get to know with other
quantum-mechanical problems, that also get reduced to the solution
of quantum pendulum.

Note that the parameter $\omega$, which is connected with the
modulation depth $\Delta l$, has (like any other parameter) a
certain small error $\delta \omega$, which during the time of one
passage $t\sim 2\pi /\omega$, leads to an insignificant correction
in the phase $2\pi(\delta\omega/\omega)$. But during the time
$t\sim\Delta T$, there occur a great number of oscillations (
phase incursion takes place) and, in the case $\Delta T\gg \tau$,
a small error $\delta \omega$ brings to the uncertainty of the
phase $\sim \Delta T \delta \omega$ which may have order $2\pi$.
Then we say that the phase is self-chaotized.

Let us introduce the density matrix averaged over a small
dispersion $\delta \omega$:
\begin{eqnarray} &\rho^{+-}_n(t)=\left(\begin{array}{cc}
W^+_n(t)&iF_n(t)\\ -iF^\ast_n(t)&W^-_n(t)\\
\end{array}\right),\,&
\end{eqnarray}
where $W^\pm_n(t)=\overline{|C^\pm_n(t)|^2},~~
F_n(t)=|\overline{C^+_n(t)C^{-\ast}_n(t)}|$. The overline denotes
the averaging over a small dispersion $\delta \omega$ \be
\overline{A(\omega,t)}=\frac{1}{2\delta
\omega}\int\limits^{\omega+\delta \omega}_{\omega-\delta
\omega}A(x,t)dx \ee

To solve (64) we can write that \be
W^+_n(t)=\overline{\sin^2\omega
t},~~W^-_n(t)=\overline{\cos^2\omega
t},~~F_n(t)=\frac{1}{2}|\overline {\sin2\omega t}|.\ee

After a simple integration of the averaging (64), for the matrix
element (65) we obtain $$W^\pm_n(t)=\frac{1}{2}(1\mp f(2\delta
\omega t)\cos 2\omega t),$$ \be
F_n(t)=F^\ast_n(t)=\frac{1}{2}f(2\delta \omega t)\sin 2\omega t,
 \ee
$$f(2\delta \omega t)=\frac{\sin2\delta \omega t}{2\delta \omega
t}.$$

At small values of time
$t\ll\overline{\tau}~(\overline{\tau}=2\pi/\delta \omega)$,
insufficient for self-chaotization  $(f(2\delta \omega t)\approx
1)$, we obtain $$W^+_n(t \ll\overline{\tau})=\sin^2 \omega
t,~~W^-_n(t \ll\overline{\tau})=\cos^2\omega
t,~~F_n(t\ll\overline{\tau})=\frac{1}{2}\sin2\omega t.$$

Comparing these values with the initial values (65) of the density
matrix elements, we see that the averaging procedure (64), as
expected, does not affect them. Thus, for small times we have
\begin{eqnarray}
&\rho^{+-}_n(t\ll\overline{\tau})=\left(\begin{array}{cc}
\sin^2\omega t&\frac{i}{2}\sin 2\omega t\\ \frac{-i}{2}\sin2\omega
t&\cos^2\omega t\\
\end{array}\right).\,&
\end{eqnarray}

One can easily verify that matrix (67) satisfies the condition
$\rho(t\ll\overline{\tau})=\rho(t\ll\overline{\tau})$, which is a
necessary and sufficient condition for the density matrix of the
pure state.

For times even smaller than $t\ll \tau\ll \overline{\tau}$, when
passages between degenerate states practically fail to occur, by
taking the limit $\omega t\ll 1$ in (67), we obtain the following
relation for the density matrix: \be
\rho^{+-}_n(t=0)=\rho^{+-}_n(t\ll\tau)=\left(\begin{array}{cc}
0&0\\ 0&1
\end{array}\right).\,
\ee This relation corresponds to the initial condition (59) when
the system is in the eigenstate $\psi^-_{2n}$. Let us now
investigate the behavior of the system at times
$t\geq\overline{\tau}$ when the system gets self-chaotized.

On relatively large time intervals $t\geq \overline{\tau}$, in
which the self-chaotization of phases  takes place, for the matrix
elements we should use general expressions (66). The substitution
of these expressions for the matrix elements (66) into the density
matrix (63) gives
\begin{eqnarray} &\rho^{+-}_n(t)=\frac{1}{2}\left(\begin{array}{cc}
1-f(2\delta \omega t)\cos 2\omega t&if(2\delta \omega
t)\sin2\omega t\\ -if(2\delta \omega t)\sin2\omega t&1+f(2\delta
\omega t)\cos2\omega t\\
\end{array}\right).\,&
\end{eqnarray}
Hence, for times $t\geq\overline{\tau}$ during which the phases
get completely chaotized, after passing to the limit $\delta
\omega t\gg1$ in (69), we obtain
\begin{eqnarray}
&\rho^{+-}_n(t\gg\overline{\tau})=\frac{1}{2}\left(\begin{array}{cc}
1-O(\varepsilon)&iO(\varepsilon)\\-iO(\varepsilon)&1+O(\varepsilon)\\
\end{array}\right),\,& \end{eqnarray}
where $O(\varepsilon)$ is an infinitesimal value of order
$\epsilon=\frac{1}{2\delta \omega t}$.

The state described by the density matrix (70) is a mixture of two
quantum states $\psi^+_{2n}$ and $\psi^-_{2n}$ with equal weights.
The comparison of the corresponding matrix elements of matrices
(70) and (69) shows that they differ in the terms that play the
role of quickly changing fluctuations. When the limit is
$t\gg\overline{\tau}$, fluctuations decrease as
$\sim\frac{1}{2\delta \omega t}$ (see Fig. 11 and 12).

Thus the system, which at the time moment $t=0$ was in the pure
state with the wave function $\psi^-_{2n}$ (68), gets
self-chaotized with a lapse of time $t\gg\overline{\tau}$ and
passes to the mixed state (70).

\begin{figure}[t]
  \centering
  \includegraphics[width=8cm]{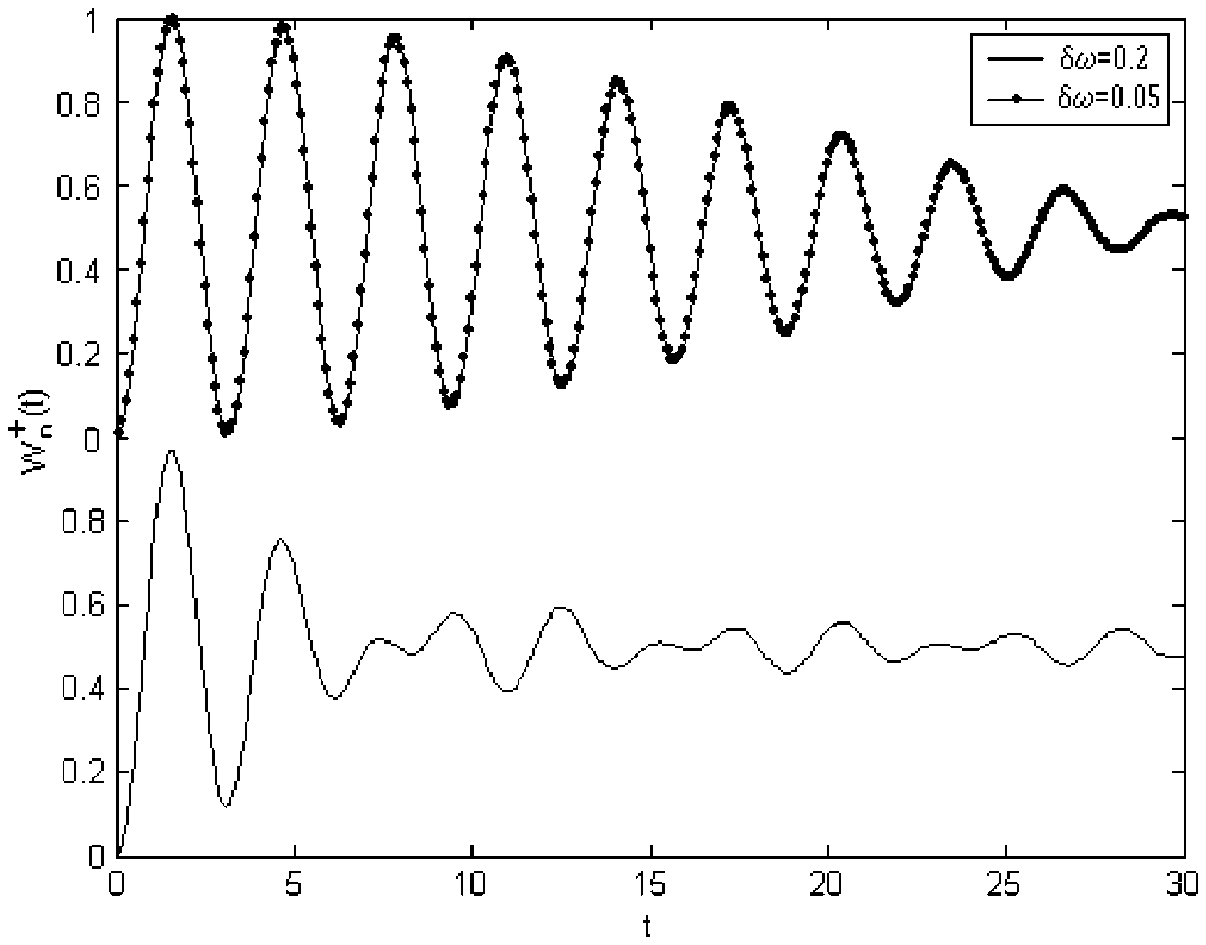}
  \caption{Time-dependence of the diagonal matrix element $W_n^+(t)$ of the
  density matrix (63), constructed by means of formulas (63),
  (66) for the parameter values $\omega=1/\tau=1,~C_n^+(0)=1,~C_n^-(0)=0.$
  As clearly seen from the Figure, the higher the
  dispersion value of the parameter $\delta \omega$, the sooner the stationary
  value $W_n^+(t>\bar{\tau}\sim\frac{1}{\delta \omega})=\frac{1}{2}$ is achieved.}\label{fig:Fig.11}
\end{figure}

In other words, at the initial moment the system had a certain
definite "order" expressed in the form of the density matrix
$\rho^{+-}_n(0)$ (68). With a lapse of time the system got
self-chaotized and the fluctuation terms appeared in the density
matrix (69). For large times $t\gg\overline{\tau}$ a new "order"
looking like a macroscopic order is formed, which is defined by
matrix (70).

After a half-period the system passes to the area  of
nondegenerate states $G$ (68). In passing through the branch
point, there arise nonzero probabilities for passages both to the
state $ce_{2n}$ and to the state $se_{2n}$. Both states
$\psi^+_{2n}$ and $\psi^-_{2n}$ will contribute to the probability
that the system will pass to either of the states $ce_{2n}$ and
$se_{2n}$. For the total probability of passage to the states
$ce_{2n}$ and $se_{2n}$ we obtain respectively

\begin{figure}[t]
  \centering
  \includegraphics[width=8cm]{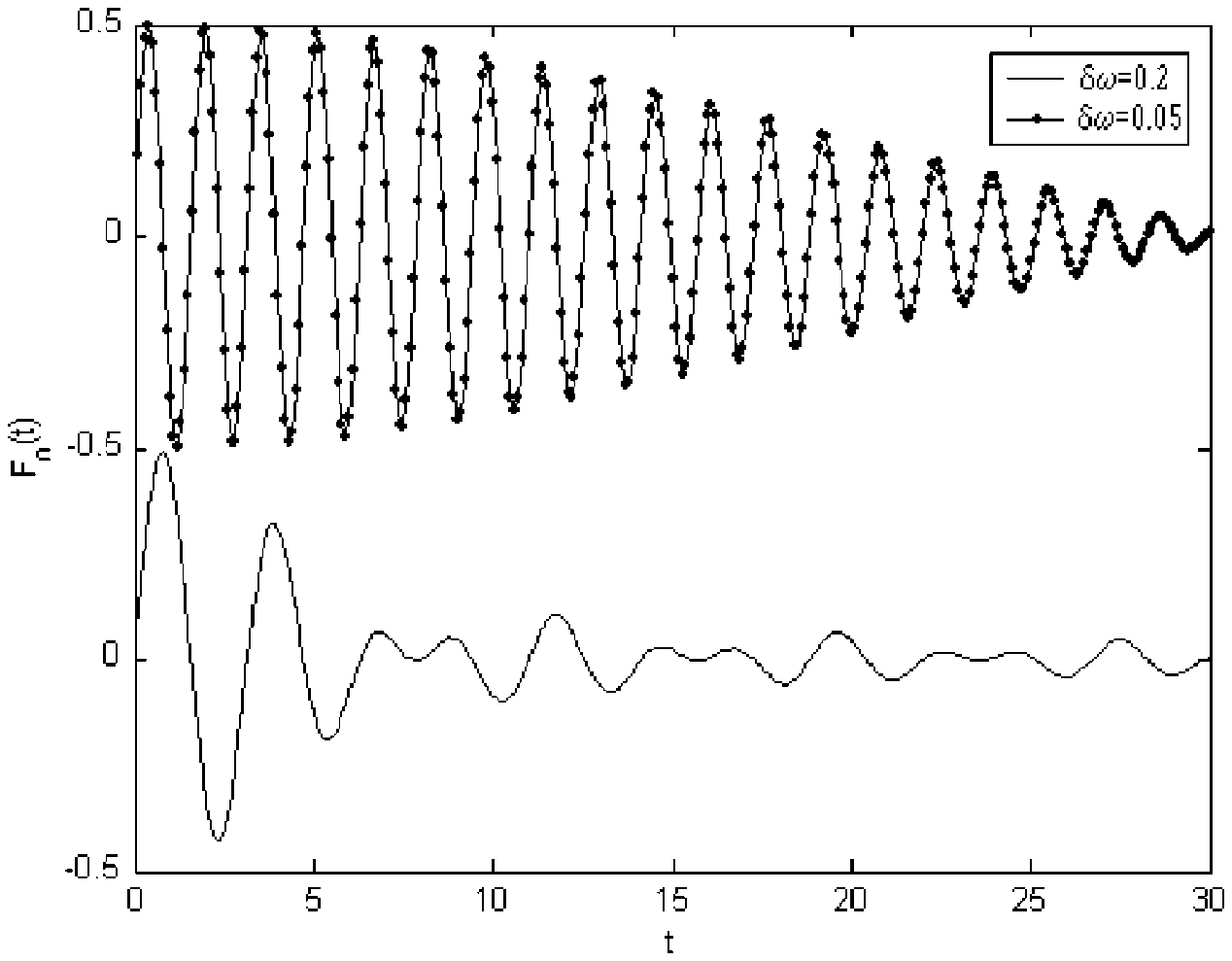}
  \caption{The vanishing of nondiagonal matrix elements of
  the density matrix (63) with a lapse of time $t>\bar{\tau}$  while
  the system remained in the degenerate area $G_-$. The graph
  is constructed for the parameter values  $\omega=1/\tau=1,~C_n^+(0)=1,~C_n^-(0)=0.$,
  with the aid of formulas (63) and (66).}\label{fig:Fig.12}
\end{figure}

$$P(\rho^{+-}_{2n}(t\gg \tau)\rightarrow
ce_{2n})=\frac{1}{2}\bigl|\frac{1}{\pi}\int\limits_{o}^{2\pi}\psi^+_{2n}
(\varphi)ce_{2n}(\varphi)d\varphi\bigr|^2+$$
$$+\frac{1}{2}\bigl|\frac{1}{\pi}\int\limits_{o}^{2\pi}\psi^-_{2n}(\varphi)
ce_{2n}(\varphi)d\varphi\bigr|^2=\frac{1}{2}\cdot \frac{1}{2}+
\frac{1}{2}\cdot \frac{1}{2}=\frac{1}{2},$$
$$P(\rho^{+-}_{2n}(t\gg \tau)\rightarrow
se_{2n})=\frac{1}{2}\bigl|\frac{1}{\pi}\int\limits_{o}^{2\pi}\psi^+_{2n}
(\varphi)se_{2n}(\varphi)d\varphi\bigr|^2+$$
\be
+\frac{1}{2}\bigl|\frac{1}{\pi}\int\limits_{o}^{2\pi}\psi^-_{2n}(\varphi)
se_{2n}(\varphi)d\varphi\bigr|^2=\frac{1}{2}\cdot \frac{1}{2}+
\frac{1}{2}\cdot \frac{1}{2}=\frac{1}{2}. \ee Thus, in the
nondegenerate area the mixed state is formed, which is defined by
the density matrix
\begin{eqnarray}
&\rho^{ik}_{2n}(t\sim\frac{T}{2}\gg\tau)=\frac{1}{2}\left(\begin{array}{cc}
1&0\\ 0&1\\
\end{array}\right),\,&
\end{eqnarray}
where $i$ and $k$ number two levels that correspond to the states
$ce_{2n}$ and $se_{2n}$.

As follows from (72), at this evolution stage of the system, the
populations of two nondegenerate levels get equalized. It should
be noted that though the direct passage (37) between the
nondegenerate levels is not prohibited, perturbation (36)
essentially influences "indirect" passages. Under "indirect"
passages we understand a sequence of events consisting a passage
$G\rightarrow G_-$ through the branch point, a set of passages
between degenerate states in the area $G_-$, and the reverse
passage through the branch point $G_-\rightarrow G$. The
"indirect" passages ocurring during the modulation half period
$T/2$ result in the equalization (saturation) of two nondegenerate
levels.

As to the nondegenerate area, the role of perturbation
$\hat{H}^\prime(t)$ in it reduces to the displacement of the
system from the left branch point to the right one.

It is easy to verify that after states (72) pass to the states of
the degenerate area  $G_+$, we obtain the mixed state which
involves four states $\xi^\pm_{2n}(\varphi)$ and
$\zeta^\pm_{2n+1}(\varphi).$

Let us now calculate the probability of four passages from the
mixed state $\rho^{ik}_{2n}$ (70) to the states
$\xi^\pm_{2n}(\varphi)$ and $\zeta^\pm_{2n-1}(\varphi)$:
$$P(\rho^{ik}_{2n}\rightarrow
\xi^\pm_{2n})=\frac{1}{2}\bigl|\frac{1}{\pi}
\int\limits_{o}^{2\pi}(ce_{2n}(\varphi)+se_{2n}(\varphi))\xi^\pm_{2n}(\varphi)
d\varphi\bigr|^2=\frac{1}{4},$$ \be P(\rho^{ik}_{2n}\rightarrow
\zeta^\pm_{2n-1})=\frac{1}{2}\bigl|\frac{1}{\pi}
\int\limits_{o}^{2\pi}(ce_{2n}(\varphi)+se_{2n}(\varphi))\zeta^\pm_{2n-1}(\varphi)
d\varphi\bigr|^2=\frac{1}{4}.\ee

As a result of these passages, in the area $G_+$ we obtain the
mixed state described by the four-dimensional density matrix
\begin{eqnarray}
&\rho^{+-}_{2n,2n+1}(t\sim
T\gg\tau)=\frac{1}{4}\left(\begin{array}{cccc} 1&0&0&0\\ 0&1&0&0\\
0&0&1&0\\ 0&0&0&1\\
\end{array}\right),\,&
\end{eqnarray}
where the indices of the density matrix (26) show that the
respective matrix elements are taken with respect to the wave
functions $\xi^\pm_{2n}(\varphi)$ and $\zeta^\pm_{2n+1}(\varphi)$
of degenerate states of the area  $G_+$.

It is easy to foresee a further evolution course of the system. At
each passage through the branch point, the probability that an
energy level will get populated is equally divided between
branched states. We can see the following regularity of the
evolution of populations for the next time periods.

After odd half periods, the population of any $n-th$ nondegenerate
level is defined as an arithmetic mean of its population and the
population of the nearest upper level, while after even half
periods as an arithmetic mean of its population and the nearest
lower level. This population evolution rule can be represented
both in the form of Table 2 and in the form of recurrent relations
$$P[n,2k]=P[n+1,2k]=\frac{1}{2}(P[n,2k-1]+P[n+1,2k-1]),$$

\begin{eqnarray}
P[n+1,2k+1]=P[n+2,2k+1]=\frac{1}{2}(P[n+1,2k]+P[n+2,2k]),
\end{eqnarray}
where $P[n,k]$ is the population value of the $n-th$ level after
time $k\frac{T}{2}$, where $k$ is an integer number. The creeping
of populations among nondegenerate levels is illustrated in
Fig.10. This Table is a logical extrapolation of the analytical
results obtained in this subsection. It shows how the population
concentrated initially on one level $n_0$  gradually spreads to
other levels. It is assumed that the extreme upper level $n_0+4$
and the extreme lower level $n_0-5$  are forbidden by condition
(34) and do not participate in the process. The results of
numerical calculations by means of formulas (75) are given in Fig.
13 and Fig. 14. Fig. 13 shows the distribution of populations of
levels after a long time $t\gg T$ when the population creeping
occurs among levels, the number of which is not restricted by
(34). Let us assume that at the initial time moment $t=0$, only
one $n_0-th$ level is populated with probability  $P(n_0)=1$.
According to the recurrent relations (75), with a lapse of each
period $T$ "indirect" passages will result in the redistribution
of populations among the neighboring levels so that, after a lapse
of time  $t=k\tau\gg T$, populations of the extreme levels will
decrease according to the law which follows from (54)$$P(n_0\pm
k)\sim\frac{1}{2^k}$$

If the number $N$ of levels defined by condition (34) is finite,
then, after a lapse of a long time, passages will result in a
stationary state in which all $N$ levels are populated with the
same probability equal to $1/N$ (see Fig. 14). Let us summarize
the results we have obtained above using the notions of
statistical physics. After a lapse of time $\Delta T$, that can be
called the time of initial chaotization, the investigated closed
system (quantum pendulum + variable field) can be considered as a
statistical system.

\begin{figure}[t]
  \centering
  \includegraphics[width=8cm]{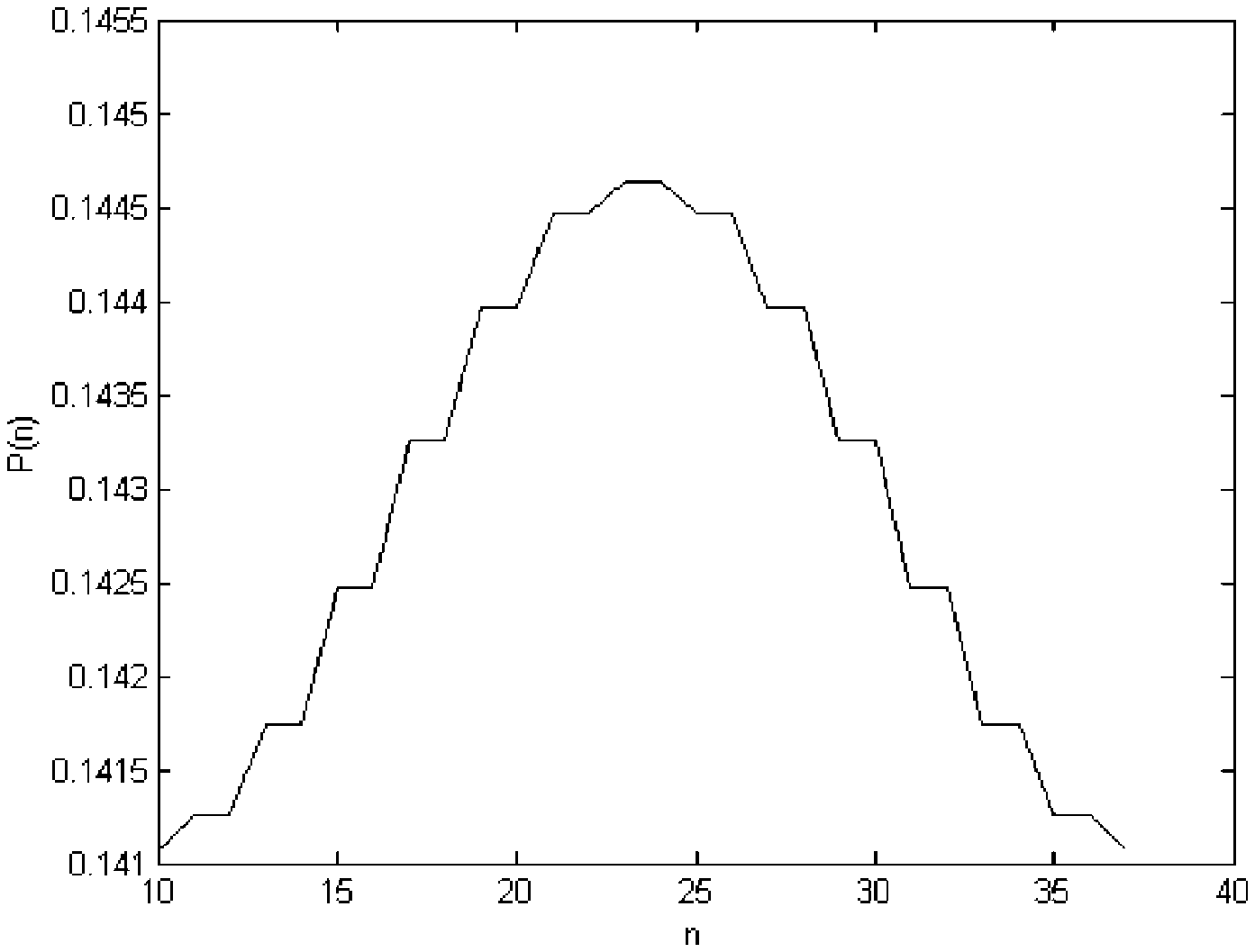}
  \caption{Results of numerical calculations performed by means of
recurrent relations (75).}\label{fig:Fig.13}
\end{figure}
\begin{table}[t]
  \centering
 \includegraphics[width=16cm]{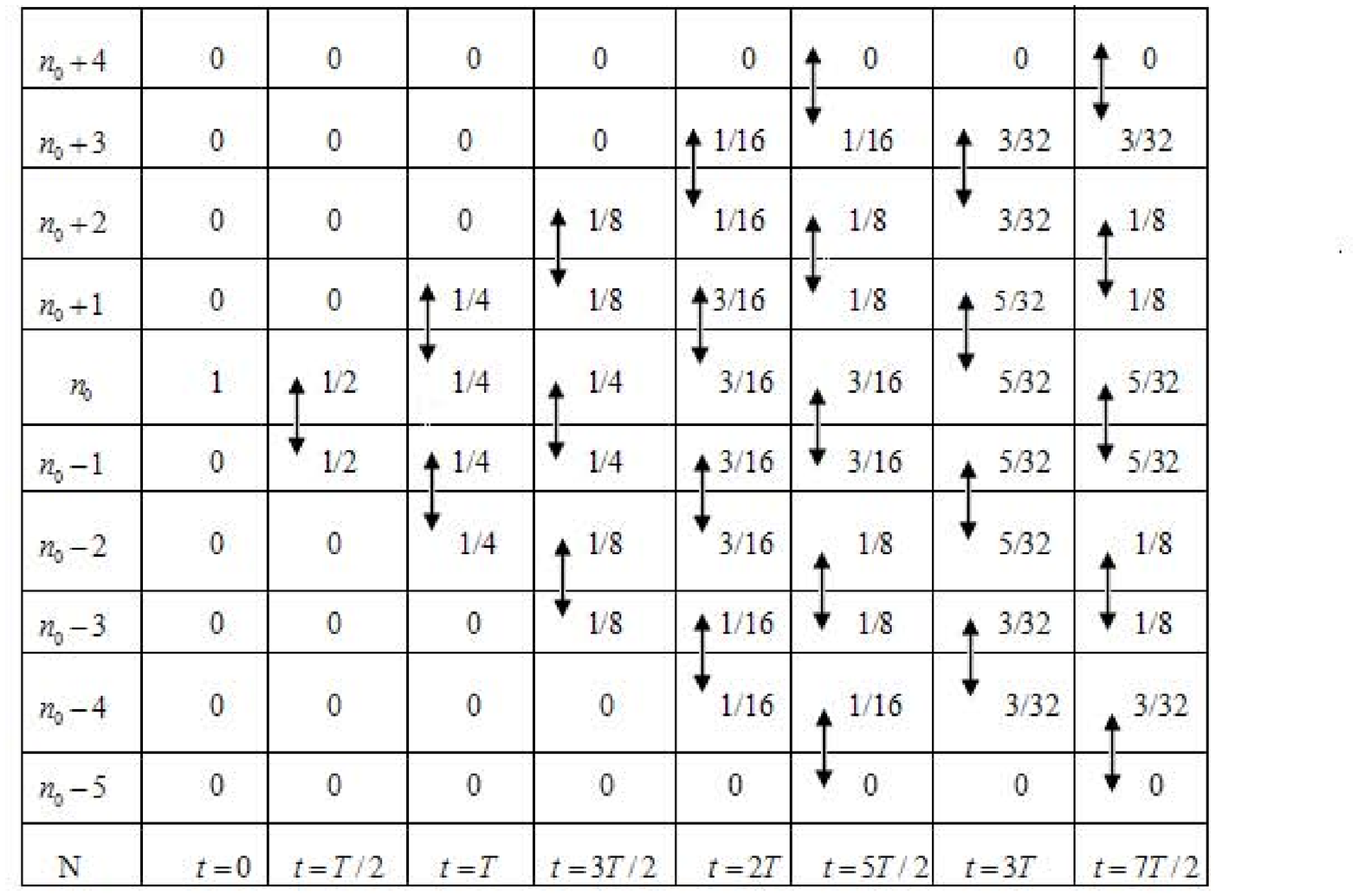}
  \caption{Evolution of populations of non-degenerate levels}\label{tab:Tab.2}
\end{table}
Formation of statistical distribution of populations of levels
$P(n)$ with a lapse of a large evolution time $t\approx1000T$ of
the system. The result shown in this figure corresponds to the
case for which the level population creeping is not restricted by
condition (34).

At that, the closed system consists of two subsystems: the
classical variable field (36) that plays the role of "a
thermostat" with an infinitely high temperature and the quantum
pendulum (7). A weak (indirect) interaction of the subsystems
produces passages between non-degenerate levels. After a lapse of
time $t\gg T$  this interaction ends in a statistical equilibrium
between the subsystems. As a result, the quantum pendulum
subsystem acquires the thermostat temperature, which in turn leads
to the equalization of level populations. The equalization of
populations usually called the saturation of passages can be
interpreted as the acquisition of an infinite temperature by the
quantum pendulum subsystem.

\begin{figure}[t]
  \centering
  \includegraphics[width=8cm]{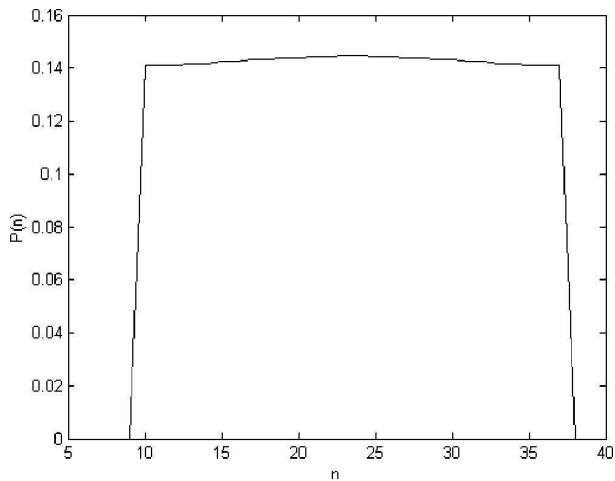}
  \caption{ Results of numerical calculations performed by means of
  recurrent relations (75).}\label{fig:Fig.14}
\end{figure}

With a lapse of a large time interval $t\approx 1000T$ the
formation of stationary distribution of populations among levels
takes place. By computer calculations it was found that in the
stationary state all $N$ levels satisfying condition (34) were
populated with equal probability $1/N$.

\subsection{Analogy between the classical and the quantum
consideration.}

The quantum-mechanical investigation of the universal Hamiltonian
(mathematical pendulum), which is reduced to the investigation of
the Mathieu-Schrodinger equation, showed that on the plane
$(E,2l)$ there exist three areas $G_+,~G_-$ and $G$ (see Fig. 8
and 9) differing from each other in their quantum properties.
Motion in the area of degenerate states $G_-$ is a quantum analog
of rotating motion of the pendulum, while motion in the area of
degenerate states $G_+$ is an analog of oscillatory motion of the
pendulum. The area $G$ lying between $G_-$ and $G_+$ can be
regarded as a quantum analog of the classical separatrix. The main
quantum peculiarity of the universal Hamiltonian is the appearance
of branching and merging points along energy term lines. Branching
and merging points define the boundaries between the degenerate
areas $G_{\pm}$ and the non-degenerate area $G$. If the system
defined by the universal Hamiltonian is perturbed by a slowly
changing periodic field, then on the plane $(E,2l)$ the influence
of this field produces the motion of the system along the Mathieu
characteristics. If, moreover, the system is in degenerate areas
for a sufficiently long time, then the phase incursion of wave
function phases occurs while the system passes through branching
points, which leads to the transition from the pure state to the
mixed one. As a result of a multiple passage through branching
points, the populations creep by energy terms (Fig. 10). The thus
obtained mixed state can be regarded as a quantum analog of the
classical stochastic layer. The number of levels affected by the
irreversible creeping process is defined by the amplitude of the
slowly changing field.

The classical mathematical pendulum may have two oscillation modes
(rotational and oscillatory), which on the phase plane are
separated by the separatrix (see Fig. 15 a).

\begin{figure}[t]
  \centering
  \includegraphics[width=16cm]{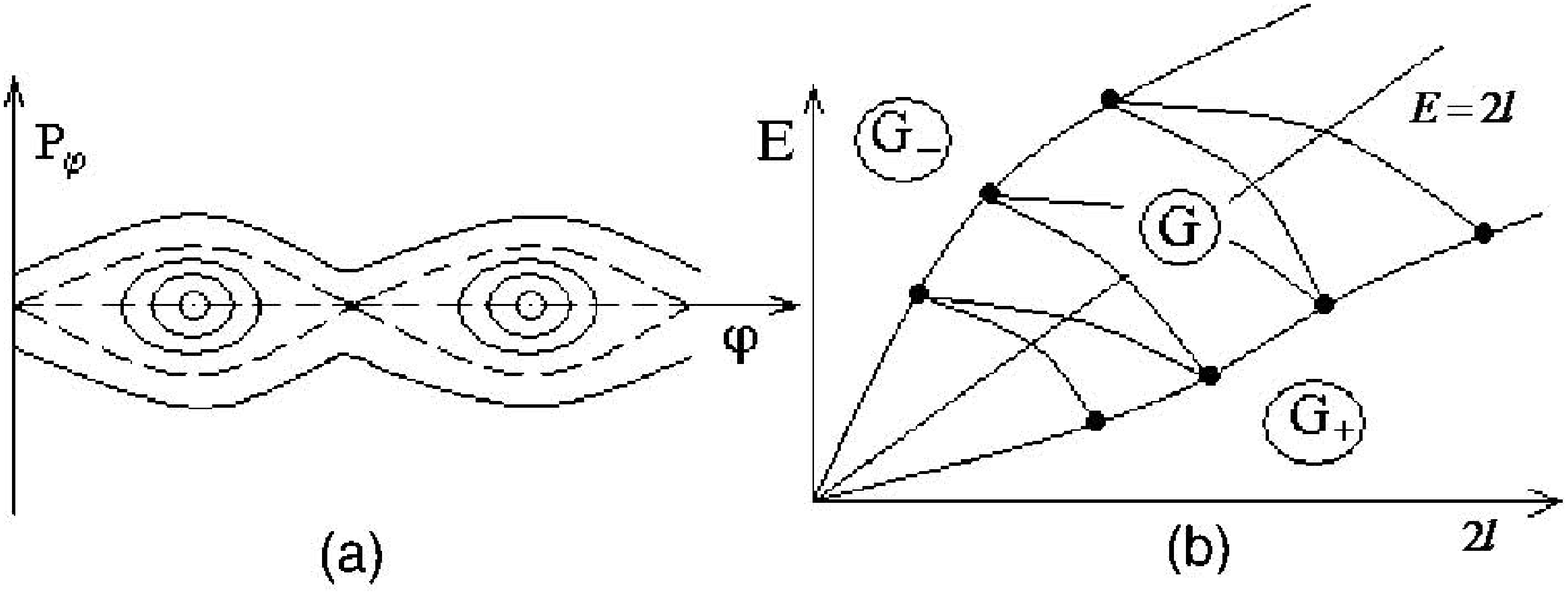}
  \caption{Analogy between the classical and quantum considerations.
  Unperturbed motion. a) Classical case. Phase plane. Separatrix;
  b) Quantum case. Specific dependence of the energy spectrum on the parameter
  (Mathieu characteristics). Degenerate $G_{\pm}$ and non-degenerate
$G$ areas of the spectrum.}\label{fig:Fig.15}
\end{figure}

On the plane $(E,2l)$ the quantum pendulum has two areas of
degenerate states $G_-$ and $G_+$. Quantum states from the area
$G_-$ possesses translational symmetry in the pendulum phase
space. These states are analogous to the classical rotational
mode. Quantum states from the degenerate area $G_+$  possess
symmetry with respect to the equilibrium state of the pendulum
(see axis in Fig. 1) and therefore are analogous to the classical
oscillatory state. On the plane $(E,2l)$, the area of
non-degenerate states $G$, which lies between the areas $G_+$ and
$G_-$, contains the line $E=2l$ corresponding to the classical
separatrix (see Fig. 15 b). If the classical pendulum is subjected
to harmonically changing force that perturbs a trajectory near to
the separatrix, then the perturbed trajectory acquires such a
degree of complexity that it can be assumed to be a random one.
Therefore we say that a stochastic motion layer (so-called
stochastic layer) is formed in the neighborhood of the separatrix
(see Fig.16 a). In the case of quantum consideration, the periodic
perturbation (36) brings about passages between degenerate states.
As a result of repeated passages, before passing to the area $G$
the system gets self-chaotized, passes from the pure state to the
mixed one and further evolves irreversibly. While it repeatedly
passes through the branch points, the redistribution of
populations by the energy spectrum takes place. Only the levels
whose branch points satisfy condition (34), participate in the
redistribution of populations (see Fig.16 b).

\begin{figure}[t]
  \centering
  \includegraphics[width=16cm]{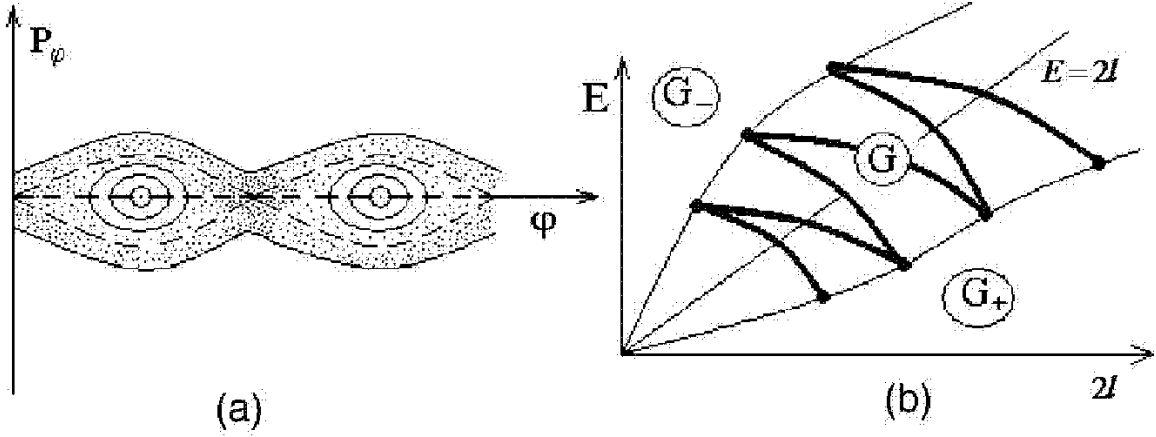}
  \caption{Analogy between the classical and quantum considerations.
Perturbed motion: a) Classical case. Stochastic trajectories in
the neighborhood of the separatrix form the stochastic layer
(cross-hatched area); b) Quantum case. The mixed state was formed
as a result of population of non-degenerate levels situated on
both sides of the classical separatrix.}\label{fig:Fig.16}
\end{figure}

\subsection{Investigation of quantum chaos of internal rotation
motion in polyatomic molecules.}

Let us consider two limiting cases of a low and a high energy
barrier. In the limit of a low barrier $V_o\rightarrow 0$, the
Mathieu-Schrodinger equation (18) implies the equation for free
rotation
 $$\frac{d^2\psi}{d\varphi
^2}+\frac{2I}{\hbar ^2}\varepsilon_r\psi=0.$$

From which for the energy spectrum we obtain $\varepsilon _r
=(\hbar ^2/2I)r^2$, where $r=0,1,2,\dots$ are integer numbers.
Using the above-given numerical estimates for the molecule of
ethane $C_2H_6$, we obtain $\varepsilon _r\approx 0.21\cdot
10^{-21}r^2J$, which corresponds to the cyclic frequency of
rotation $\approx 2.0\cdot10^{12}r^2rad/s$. Comparing the
expression for energy with the value of the ethane molecule
barrier we see that only levels with a sufficiently large quantum
number $(r>10)$ are located high above the barrier and it is only
for such levels that the considered limit is valid.

In another limiting case of a high barrier $V_o$ the rotator is
most of the time inside one of the potential wells where it
performs torsional motions. In that case $\alpha$ can be treated
as a small angle. After expanding the potential energy in the
Schrodinger equation (9) into small angles $\cos 2\alpha \approx
1-2\alpha ^2$, we obtain a quantum equation for the oscillator,
whose energy spectrum has the form $\varepsilon_r=(r+1/2)\hbar
\omega$, where $\omega=n\sqrt{V_o/2I}\approx 6.0\cdot
10^{13}rad/s$. For the energy spectrum of small torsional
oscillations we obtain $\varepsilon_r\approx(r+1/2)\cdot 0.63\cdot
10^{-20}J$. If we compare the obtained expression for the spectrun
with the corresponding numerical value of the barrier, then we can
see that only the first two levels $\varepsilon_o\approx 0.32\cdot
10^{-20}J,$ $\varepsilon_1\approx 0.95\cdot 10^{-20}J$ and
$\varepsilon_2\approx 1.58\cdot 10^{-20}J$ are located in the well
but not at a sufficiently large depth that would allow us to
assume that passages between them correspond to small
oscillations. Thus we can conclude that for  internal rotation of
the molecule of ethane $C_2H_6$ the approximation of small
oscillations is not carried out sufficiently well, while the
approximation of free rotation is carried out for large quantum
numbers.

A real quantitative picture of the internal rotation spectrum can
be obtained by means of the Mathieu-characteristics when the
points of intersection of the line $l=l_o$ with the Mathieu-
characteristics is projected on to the energy axis (see Fig.17).

These conclusions are in good agreement with experimental data. In
particular, in the experiment we observed the infrared  absorption
by molecules of $C_2H_6$ at a frequency $\sim8.7\cdot 10^{12}Hz$
\cite{Shimanouchi}. For an energy difference between the levels
participating in the absorption process we have the estimate
$\Delta \varepsilon_{exper}\sim 0.54\cdot 10^{-20}J$. Comparing
the experimental result with the energy difference between two
neighboring levels in the case of approximation of small
oscillations, we obtain $$\Delta \varepsilon_o=2\pi\nu_o\hbar,$$
where $\nu_o=\frac{3}{2\pi}\sqrt{\frac{V_o}{2I}}$ is the frequency
of small oscillations.

Inserting the parameters values for molecules of $C_2H_6$, we
obtain the estimate $\Delta \varepsilon_o= 0.63\cdot 10^{-20}J$.

\begin{figure}[t]
  \centering
  \includegraphics[width=16cm]{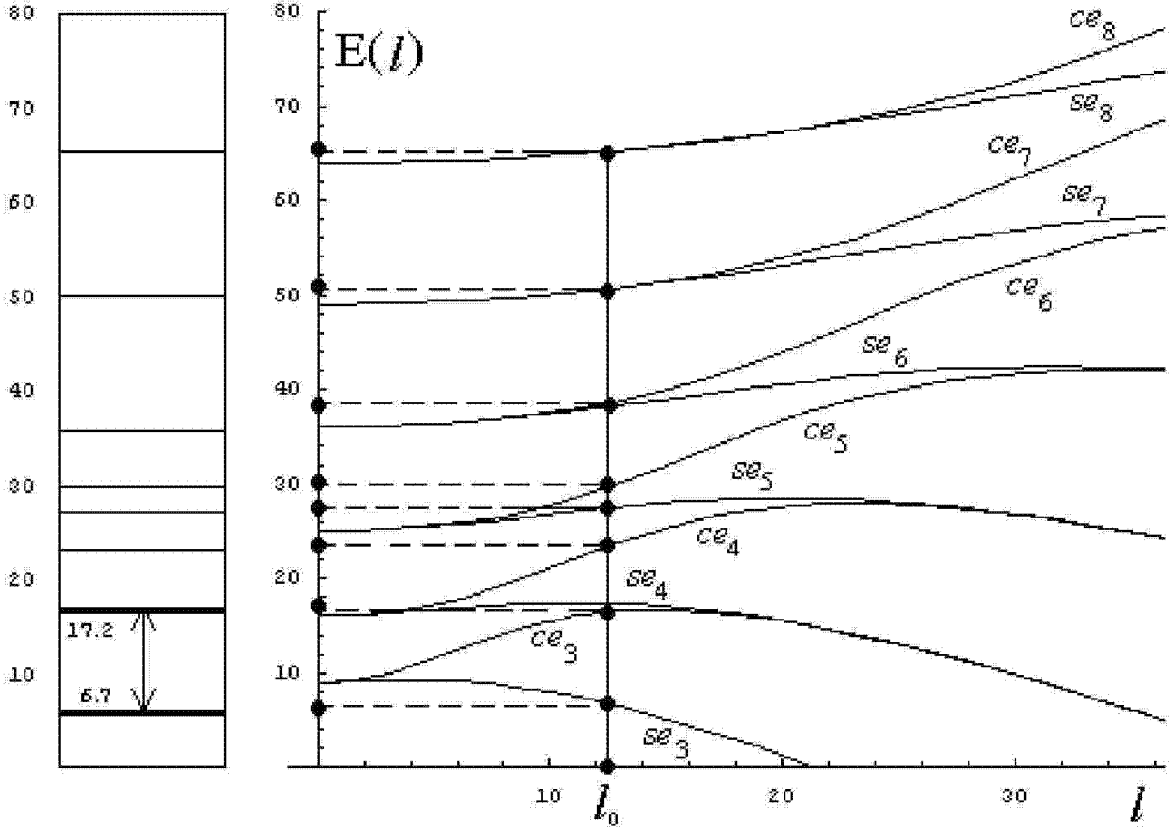}
  \caption{The graphic method of finding energy terms of internal
rotation. For the ethane molecule
$l_0=\frac{2I}{9\hbar^{2}}V_{0}$.} \label{fig:Fig.17}
\end{figure}

After comparing the obtained estimates with the energy difference
between the states described by the wave functions
$ce_3(\varphi,l_o), se_3(\varphi, l_o)$ and applying formula (20),
we obtain $\Delta \varepsilon
(ce_3(\varphi,l_o)\leftarrow\rightarrow se_3(\varphi, l_o))\sim
0.54\cdot10^{-20}J$.

Comparative analysis of the obtained estimates gives us the
grounds to conclude that the energy levels corresponding to the
states $ce_3(\varphi,l_o), se_3(\varphi, l_o)$ participate in the
infrared absorption revealed in the experiment.

As we see from Fig.17 energy levels involved in the process are
distributed in a random manner. In some sense, this resembles RMT.
However as number of levels is small, and energy spectrum is well
defined.

Let us assume that the considered quantum system is subjected to a
radio frequency (RF) monochromatic pumping whose frequency
$\Omega$ satisfies the condition $\Omega \ll V_o / \hbar$. This
causes a slow modulation of quick electron motions in a molecule.
The formation of an energy barrier $V_o$ is a result of the
averaging over quick electron motions and thus it is obvious that
due to the pumping effect the barrier value is time-dependent, \be
V_o\rightarrow V_o+\Delta V\cos \Omega t.\ee The depth of
modulation $\Delta V$ depends on a pumping power. By replacing
(76) we obtain the time-dependent Hamiltonian
$$\hat{H}=\hat{H}_o(\varphi)+\hat{H}^\prime(\varphi,t),$$
\be \hat{H}_o(\varphi )=-\frac{\partial ^2}{\partial
\varphi^2}+l_o\cos 2\varphi, \ee
$$\hat{H}^\prime (\varphi,t)=\Delta l\cos 2\varphi \cos \Omega
t.$$
\be
\Delta l=\frac{2I}{n^2\hbar ^2}\Delta V.\ee

Simple calculations show that the matrix elements of perturbation
$\hat{H}^\prime (\varphi ,t)$ with respect to the wave functions
of the nondegenerate area $G$ are equal to zero \be \langle
ce_n|\hat{H}^\prime(\varphi ,t)|se_n\rangle \sim\Delta
l\int\limits^{2\pi}_o ce_n(\varphi)\cos2\varphi
se_n(\varphi)d\varphi =0,\ee where $n$ is any integer number.
Therefore perturbation (78) cannot bring about passages between
nondegenerate levels.

The interaction $\hat{H}^\prime (\varphi ,t)$, not producing
passages between levels, should be inserted in the unperturbed
part of the Hamiltonian. The Hamiltonian obtained in this manner
can be considered as slowly depending on time.

Thus, in the nondegenerate area the Hamiltonian can be written in
the form
\be
\hat{H}=-\frac{\partial ^2}{\partial \varphi ^2}+l(t)\cos
2\varphi,\ee

Because of the modulation of the parameter $l(t)$ the system
passes from one area to another, getting over the branch points.

As different from the nondegenerate state area $G$, in the areas
of degenerate states $G_-$ and $G_+$, the nondiagonal matrix
elements of perturbation $\hat{H}^\prime(t)$  (78) are not equal
to zero. For example, if we take the matrix elements with respect
to the wave functions $\psi^\pm_{2n+1}$ , then for the left
degenerate area $G_-$ it can be shown that \be
H^\prime_{+-}=H^\prime_{-+}=\langle
\psi^+_{2n+1}|\hat{H}^\prime(\varphi ,t)|\psi^-_{2n+1}\rangle
\sim\Delta l
\int\limits_{o}^{2\pi}\psi^+_{2n+1}\psi^{-\ast}_{2n+1}\cos
2\varphi d\varphi \not= 0. \ee Note that the value $H^\prime_{+-}$
has order equal to the pumping modulation (78) depth $\Delta l.$

Analogously to (81), we can write an expression for even $2n$
states as well.

An explicit dependence of $\hat{H}^\prime(\varphi, t)$ on time
given by the factor $\cos \Omega t$ is assumed to be slow as
compared with the period of passages between degenerate states
that are produced by the nondiagonal matrix elements
$H^\prime_{+-}$. Therefore below the perturbation
$\hat{H}^\prime_{+-}(\varphi ,t)$ will be assumed to be the
time-independent perturbation that can bring about passages
between degenerate states.

In a degenerate area the system may be in the time-dependent
superpositional state \be
\psi_{2n}(t)=C^+_n(t)\psi^+_{2n}+C^-_n(t)\psi^-_{2n}.\ee The
probability amplitudes $C^{\pm}_n(t)$ are defined by means of the
fundamental quantum-mechanical equation, expressing the casuality
principle. We write such equations for a pair of doubly degenerate
states: $$-i\hbar\frac{dC^+_n}{dt}=(E_{on}+H^\prime
_{++})C^+_n+H^\prime_{+-}C^-_n,$$ \be
-i\hbar\frac{dC^-_n}{dt}=H^\prime _{+-}C^+_n+(E_{on}+H^\prime
_{--})C^-_n. \ee where the matrix elements are taken with respect
to degenerate wave functions (see (81)) and $E_{on}$ is the energy
of the $n-th$ degenerate level near a branch  point.

Let us assume that at the initial moment of time the system was in
the degenerate state $\psi^-_{2n}$. Then as initial conditions we
should take \be C^-_n(0)=1,~~~ C^+_n(0)=0. \ee Having substituted
(82) into (81), for the amplitudes we obtain
$$C^+_n(t)= iexp(\frac{i}{\hbar}Et)\sin \omega t,$$
\be C^-_n(t)= exp(\frac{i}{\hbar}Et)\cos \omega t, \ee
$$E=E_{on}+H^\prime_{\pm\pm},$$ where $\omega
=\frac{2\pi}{\tau}=\frac{H^\prime _{+-}}{\hbar}$ is the frequency
of passages between degenerate states, $\tau$ is the passage time.

Note that the parameter $\omega$ has (like any other parameter) a
certain small error $\delta \omega$, which during the time of one
passage $t\sim 2\pi /\omega$, leads to an insignificant correction
in the phase $2\pi(\delta\omega/\omega)$. However, if during the
time $t\sim\Delta T$, when the system is the degenerate area
$(\Delta T<T, T=2\pi/\Omega )$ there occurs a great number of
passages $(\Delta T\gg \tau )$, then for  $ \Delta T \delta \omega
\approx 2\pi$, a small error $\delta \omega$ leads to the phase
uncertainty. Then we say that the phase is self-chaotized. The
self-chaotization formed in this manner can be regarded as the
embryo of a quantum chaos which, as we will see in the sequel,
further spreads to other states.

After a half-period, the system passes to the area  of
nondegenerate states $G$. In passing through the branch point,
there arise nonzero probabilities for passages both to the state
$ce_{2n}(\varphi)$ and to the state $se_{2n}(\varphi)$. Thus, in
the nondegenerate area the mixed state is formed, which is defined
by the density matrix

\begin{eqnarray}
&\rho^{ik}_{2n}(t\sim\frac{T}{2}\gg\tau)=\frac{1}{2}\left(\begin{array}{cc}
1&0\\ 0&1\\
\end{array}\right),\,&
\end{eqnarray}
where $i$ and $k$ number two levels that correspond to the states
$ce_{2n}(\varphi)$ and $se_{2n}(\varphi)$.

As follows from (86), at this evolution stage of the system, the
populations of two nondegenerate levels get equalized. It should
be noted that though the direct passage (79) between the
nondegenerate levels is prohibited, perturbation (78) essentially
influences "indirect" passages. Under "indirect" passages we
understand a sequence of events consisting a passage $G\rightarrow
G_-$ through the branch point, a set of passages between
degenerate states in the area $G_-$, and the reverse passage
through the branch point $G_-\rightarrow G$. The "indirect"
passages occurring during the modulation half period $T/2$ result
in the equalization (saturation) of two nondegenerate levels.

Thus "indirect" passages are directly connected with a quantum
chaos. Hence, by fixing "indirect" passages we thereby fix the
presence of a quantum chaos.

Let us assume that the investigated molecule is a component of a
substance in a gaseous or liquid state. Then the molecular thermal
motion, which tries to establish an equilibrium distribution of
populations according to Boltzman's law, will be a "competing"
process for the quantum chaos described above. Using thermodynamic
terminology, we can say that the considered quantum system is
located between two thermostats. One of them with  medium
temperature $T_o$ tries to retain thermal equilibrium in the
system, while the other, having an infinite temperature, tries to
equalize the populations.

An equation describing the change of populations according to the
scheme shown in Fig.18 has the form \be
\frac{dn_i}{dt}=-2Wn_i-\frac{n_i-n^{(o)}_i}{T_1}, \ee

\begin{figure}[t]
  \centering
  \includegraphics[width=16cm]{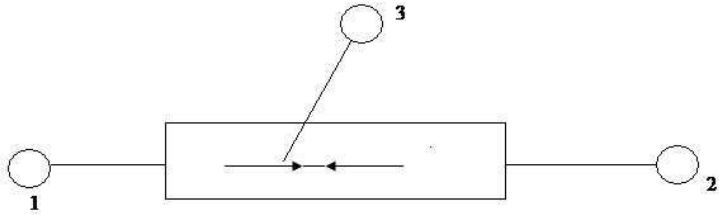}
  \caption{A thermodynamic scheme of the process. Subsystem 1 is a
usual thermostat with temperature $T_0$ , subsystem 2 is a
thermostat having an infinite temperature and consisting of the
interaction $\hat{H}'(\varphi,t)$ (78); the subsystem 3 is the
quantum system corresponding to internal rotations of  molecules
and being able to receive energy from subsystem 2 and to transfer
it to subsystem 1.}\label{fig:Fig.18}
\end{figure}
where $2W$ is a probability of "indirect" passages and $T_1$ is
the time of thermal chaotization. Bloembergen, Parcell and Pound
used equation (87) to describe the process of saturation of
nuclear magnetic resonance in solid bodies \cite{Bloembergen}. For
a stationary distribution of populations from (87) we obtain \be
n_i = n^{(o)}_i\frac{1}{1+s},\ee where $s=2WT_1$ is called the
saturation parameter. For $s\gg1$ "indirect" passages have a
stronger effect on the system than thermal processes.

Difference from the magnetic resonance consists in the expression
of the transition probability $W$. Whereas, in the case of
magnetic resonance, transition probability $2W\sim H^2_1/\Delta $
is proportional to the pumping intensity  $H_1$ and inversely
proportional to the line width $\Delta $, in the case of
non-direct transitions, transition probability $2W\sim 1/T$ does
not depend on the oscillation amplitude $\Delta V$ and is
inversely proportional to the oscillation period $T$ (i.e. to the
time of motion between branch points). But, it is worth to keep in
mind, that the amplitude of oscillation must be large enough for
the system to achieve the branch points.

Thus, along with the conditions that the perturbation is adiabatic
$\Omega\ll V_0/\hbar$, phase incursion $\Delta T\gg\tau$ and
self-chaotization during the process of multiple transitions
between the degenerated states takes place $\Delta
T\delta\omega\approx2\pi,~\delta \omega\ll\omega,~\Delta T<T$. the
condition $s\gg1$ is a necessary condition for the formation of a
quantum chaos. In the opposite limiting case $s\ll1,~n_i\approx
n_i^{(0)}$, the quantum chaos will be completely suppressed by
thermal motion. Let us proceed to discussing the quantum chaos
possible experimental observation. The most suitable material for
this purpose in our opinion is ethane  $C_2H_6$. But, by use of
easy estimations one can prove, that for the ethane in the gaseous
state, conditions of the quantum chaos observation hardly can be
achieved. Really the requirement that modulation must be slow
$\Omega\ll V_0/\hbar\approx 10^{14}sec^{-1}$, leads to the
necessity to take the modulation frequency from the transition
radio frequency range $\Omega\approx 0.3\cdot 10^{12}sec^{-1}.$

In the case of gases the thermal chaotization time is estimated by
the formula $T_1\approx\frac{d}{\bar{\nu}}$ , where $d$ is the
molecule size,
$\bar{\nu}=\sqrt{\bar{\nu^2}}=\sqrt{3/2}(2kT_0/m)^{1/2}$ is a mean
motion velocity of molecules. After substituting the numerical
values for $C_2H_6$, we obtain
$T_1\approx0.3\cdot10^{-11}\sqrt{1/T_0},~T_0\approx
200^0K>184^0K$, (where $184^0K$ is the boiling point of the ethane
at normal pressure). Then for the time of thermal chaotization and
for the saturation parameter we get respectively:
$T_1\approx0.3\cdot10^{-12}sec,~S\sim(T_1/T)\sim1$. Hence, thermal
chaotization and quantum chaos are equally manifested in the
system.

In the case of a liquid under $T_1$ we should understand the mean
time of the settled life of a molecule, which is about
$10^{-8}sec$. Relatively large times of relaxation in liquids
ensure the fulfillment of the saturation condition $(s\approx
T_1/T\approx 10^4)$. In such a way, for ethane $C_2H_6$ in the
liquid state, quantum chaos caused by the non-direct transitions
has stronger influence upon the system then common thermal
chaotization. At the same time, the state with the equally
populated levels will be formed in the system.

Suppose, the system (ethane in the liquid state) is subjected to
the action of a infrared radiation field and infrared absorption
at the frequency $\sim 289 sec^{-1}$ is observed
\cite{Shimanouchi}. This absorption corresponds to the transitions
between the states $se_3\leftrightarrow ce_3$. Furthermore, let us
suppose that along with the infrared, the radio frequency pumping
which can cause the non-direct transitions is applied. This in its
turn leads to the discontinuance of the infrared absorption. This
phenomenon can be considered as an observation of quantum chaos.

Now we consider the example of the reaction of isomerization
$LiCH\rightleftarrows CHLi$, that was discussed in subsection $E$.
As was mentioned above the most simplified model, corresponding to
this process, can be presented with the aid of Mathieu-Schrodinger
equation (77). It is known, that energy barrier of transition of
the process of vibration of the atom  $Li$ in the process of
isomerization $V\simeq 7\cdot 10^{-20}J$. The moment of inertia is
$I_{Li}=1.1\cdot10^{-46}mcg^2$ That is why, in the case
$\frac{V_0}{\hbar\omega_0}=\sqrt{2V_0I_{Li}}\approx 19$ number of
vibration energy levels are packed in the potential well. Now it
is not hard to determine energy spectrum with the aid of Mathieu
characteristics.

\begin{figure}[t]
  \centering
  \includegraphics[width=16cm]{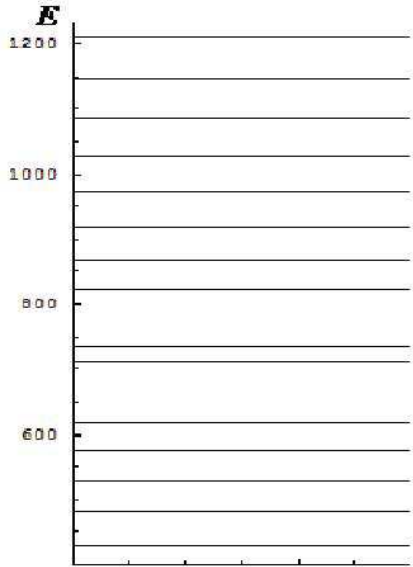}
  \caption{Energy spectrum of the process of isomerization $LiCH\rightleftarrows CHLi$ ,
  obtained as spectrum of eigenvalues of Mathieu-Schrodinger
  equation, when $l_0=\frac{I_{Li}}{2\hbar^2}V_0\simeq350.$ Speed
  of energy levels close to $E=2l_0\sim700$ can
  easily be observed in figure.}\label{fig:Fig.19}
\end{figure}

For this process the presence of chaotically distributed levels in
infrared region was established with the aid of numerical
experiments.

Now let us deal with the problem of nonlinear quantum resonance
from which we have begun review. To begin with let us estimate
resonance value of $I_0$, close to which variation of values of
$\Delta I$ may take place. By definition resonance value of action
$I_0$ is determined from equation
$\omega-\omega_0+\omega_n(I_0)=0$. By taking typical values of
parameters, know from optics:
$\omega_0\sim10^{15}sec^{-1},~\omega-\omega_0\simeq10^{12}sec^{-1},~\gamma\simeq
4\cdot 10^{40}J/kg\cdot m^4$ we obtain:
$I_0=\frac{\omega-\omega_0}{\gamma}\frac{2m\omega_0^2}{3\pi}\simeq
0.5\cdot 10^{-29}J\cdot sec$. This value of action $I/\hbar\simeq
0.5\cdot 10^5$ times as much as Plank constant. So, resonance
takes place on very high levels (in quasi-classical region) of
atoms potential well. As we have done in previous cases, it is
possible to find variation spectrum of action $\Delta I $ with the
aid of Mathieu characteristics. For this, we must know the value
of dimensionless parameter $l_0=2V(I_0)/\omega'\hbar^2$. It is
easy to find the value $\omega'\simeq2\cdot10^{41}J^{-1}sec^{-2}$
and for strong light fields $f_0\simeq10^8 V/m$. As a result we
obtain $l_0\simeq10^{40}$. This means in its turn, that there are
a large number of levels in potential well. Because of the large
value of $l_0$, it is not possible to determine spectrum by
method, used previously, with the aid of Mathieu characteristics.
However, by extrapolating it is possible to expect, that at
approaching the top of well, alongwith the common attraction of
levels, their chaotization will take place. In our opinion,
chaotically distributed levels of great density are the basis for
random matrix assumption.

\section{The Peculiarities of Energy Spectrum of Quantum Chaotic Systems. Random Matrix Theory.}

In the first part of this work we have considered the possibility
of formation of mixed state in quantum chaotic system. The
peculiarities of spectral characteristics of the universal
Hamiltonian (7) of concrete physical system corresponding to the
Mathieu-Schrodinger equation (9) was a cause of this.

In the present part we shall try to get the same results
proceeding from more general consideration. In particular, we
shall use methods of random matrixes theory
\cite{Haake,Stockman},\cite{Gaspard}.

Study of quantum reversibility and motion stability is of great
interest \cite{Hiller}. This interest is due to not only the
fundamental problem of irreversibility in quantum dynamics, but
also to practical application. In particular, it reveals itself in
relation to the field of quantum computing \cite{Turchett}.

A quantity of central importance which has been on the focus of
many studies \cite{Haug,Prosen,Beneti,Jalabert} is the so-called
fidelity $f(t)$, which measures the accuracy to which a quantum
state can be recovered by inverting, at time $t$, the dynamics
with a perturbed Hamiltonian:\be
f(t)=|<\psi|e^{i\hat{H}t}e^{-i\hat{H_{0}}t}|\psi>|^{2},\ee where
$\psi$ is the initial state which evolves in time t with the
Hamiltonian $\hat{H_{0}}$ while $\hat{H}=\hat{H_{0}}+V$ is
perturbed Hamiltonian. The analysis of this quantity has shown
that under some restrictions, the series taken from $f(t)$ is
exponential with a rate given by the classical Lyapunov exponent
\cite{Gaspard}. But here a question appears. The point is that the
origin of the dynamic stochasticity, which is a reason of
irreversibility in classical case, is directly related to the
nonlinearity of equations of motion, For classical chaotic system
this nonlinearity leads to the repulsion of phase trajectories at
a sufficiently quick rate \cite{Sagdeev,Lichtenberg}.

In case of quantum consideration, the dynamics of a system is
described by a wave function that obeys a linear equation and the
notion of a trajectory is not used at all. Hence, at first sight
it seems problematic to find out the quantum properties of
systems, whose classical consideration reveals their dynamic
stochasticity.

In this part of the paper, by using of method of random matrix
theory (RMT), we try to show that quantum chaotic dynamics is
characterized by the transition from a pure quantum-mechanical
state into the mixed one.

With this purpose we shall consider a case when the system's
Hamiltonian may be presented in the form:\be
\hat{H}(t)=\hat{H_{0}}+V(t),\ee where $\hat{H_{0}}$ is chaotic
Hamiltonian with irregular spectrum, $\hat{V}(t+T_{0})=\hat{V}(t)$
periodic in time perturbation. We shall try to show that in this
case irreversibility in the system appears as a result of loss of
information about the phase factor of the wave function. At the
same time, unlike of the second part of the paper where we have
proved this fact proceeding from the spectral peculiarities of the
concrete system, now we shall prove this in more general manner.
For this reason we shall make use of the fact that according to
RMT the eigenvalues of chaotic Hamiltonian can be considered as a
set of random number \cite{Relano,Retamosa}.

After taking (90) into account, the solution of the time-dependent
Schrodinger equation \be i\frac{\partial |\psi(t)>}{\partial
t}=\hat{H}(t)|\psi(t)>\ee can be written formally with the help of
a time-dependent exponential \cite{Haake} \be
U(t)=exp[-i\int_{0}^{t}dt^{\prime}\hat{H}(t^{\prime})],\ee where
the positive time ordering requires:\be [A(t)B(t')]_t= \left\{
\begin{array}{l}
A(t)B(t')\hskip 0.7cm{\mbox if}\hskip 0.2cm t>t' \\
B(t')A(t)\hskip 0.7cm {\mbox if} \hskip 0.2cm t<t'
\end {array} .
\right. \ee

In our case $H(t+T_{0}k)=H(t),~~k=1,2,\ldots$ the evolution
operator referring to one period $T_{0}$, the so-called Flouqet
operator $U(T_{0})=F$ \cite{Haake}, is worthy of consideration,
since it yields the stroboscopic view of the dynamics \be
|\psi(kT_{0})>=(\hat{F})^{n}|\psi(0)>.\ee The Flouqet operator
being unitary has unimodular eigenvalues. Suppose we can find
eigenvectors $|\varphi_{n}>$ of the Flouqet operator $$\hat{F}
|\varphi_{n}>= e^{-i\varphi_{n}}|\varphi_{n}>,$$ \be
<\varphi_{n}|\varphi_{m}>=\delta _{nm}.\ee Then, with the
eigenvalue problem solved, the stroboscopic dynamics may be
written explicitly \be |\psi(kT_{0})>=\sum_{n}
e^{-ik\varphi_{n}}<\varphi_{n}|\psi(0)|\varphi_{n}>.\ee As it was
mentioned above, our aim is to prove that one of the signs of the
emergence of quantum chaos is a formation of the mixed state.
Being initially in a pure quantum-mechanical state, described by
the wave function $|\psi>$, the system during the evolution makes
an irreversible transition to the mixed state.

The information about whether the system is in the mixed state or
in the pure one, may be obtained from the form of the density
matrix \cite{Feynman}. Using (96) as a density matrix of the
system we get the following expression:\be
\rho_{nm}(t=kT_{0})=A_{nm}e^{ik(\varphi_{m}-\varphi_{n})},\ee \be
A_{nm}=<\varphi_{m}|\psi(0)><\psi(0)|\varphi_{n}>. \ee

Exponential phase factors of the non-diagonal matrix elements
$\rho_{nm}(t)$ express the principle of quantum coherence
\cite{Schleich}, and correspond to the complete quantum-mechanical
description of the system in pure quantum-mechanical state. While
they are not equal to zero, the system is in the pure state. So,
to prove the formation of the mixed state one has to show zeroing
of non-diagonal elements of density matrix.

According to the main hypothesis of the random matrix theory
\cite{Haake}, the phase $f(n,m)=\varphi_{m}-\varphi_{n}$ in the
exponential factors of the non-diagonal matrix elements in (97) is
a random quantity. So, it is clear that the values of matrix
elements of density matrix of the chaotic quantum-mechanical
system $\rho_{nm}(k)$ are random values too. Taking a statistical
average of expression (97), we have \be
<\rho_{nm}(k)>=<A_{nm}e^{ik(\varphi_{m}-\varphi_{n})}>.\ee Then in
case of random phase $f(n,m)$ has a normal dispersion, we get
\cite{Skrinnikov}: \be \rho_{nm}(k)\sim
e^{-\frac{\sigma^{2}k^{2}}{2}}.\ee

This phenomenon is connected with the "phase incursion".
Uncertainty of phase in (97) is accumulated little by little with
time. Finally, when $k >\sqrt{2}/\sigma,$ the uncertainty of phase
is of order $2\pi$ , and the phase is completely chaotized. As a
result the system passes into mixed state.

According to the theory presented above, calculations were made
for concrete physical systems. In particular, in work
\cite{Skrinnikov} chaotic system of connected oscillators was
studied, and in work \cite{Kereselidze} Kepler's asymmetry
problem. The numerical results obtained in these works prove the
formula (100) to be correct.

After formation of the mixed state the quantum-mechanical
consideration loses its meaning and there is a need to use a
kinetic description.

For derivation of master equation let us split the density matrix
operator $\hat{\rho}$ on a slow  $\hat{\rho}_R$ and a fast varying
$\hat{\rho}_{NR}$ operators: \be
\hat{\rho}=\hat{\rho}_R+\hat{\rho}_{NR}. \ee Relevant part
$\hat{\rho}_{R}$ in the basis $|\phi _n\rangle$ contains only
diagonal elements, whereas nonrelevant part $ \hat{\rho}_{NR}$
contains only nondiagonal elements. These elements, as it was
shown in previous section, contain  fast oscillating exponents and
when taking average over the ensemble, the zeroing of them takes
place. Elimination of the diagonal part from the density matrix is
linear operation, which satisfies the property of projection
operator $\hat {D}^2=\hat{D}$ \cite{Skrinnikov} \be \hat{\rho}_R
=\hat{D}\hat{\rho}, ~~ \hat{\rho}_{NR}=(1-\hat{D}) \hat{\rho}. \ee

Let us note that this reflection is nonreversible. Due to the
zeroing of nondiagonal part of the density matrix a part of
information is lost.

Inasmuch as relevant statistical operator $\hat{\rho}_R(t)$ is
different from the total operator  $\hat{\rho}(t)$, generally
speaking it does not suit Liouville-Fon Neumann equation
\cite{Feynman} \be \frac{\partial \hat{\rho}}{\partial t}+i\hat{L}
\hat{\rho}=0, \ee where $\hat{L}$ is Liouville operator. After
acting on the equation (103) with $\hat{D}$ operator, we get \be
\frac{\partial \hat{\rho}_R}{\partial t}+i\hat{D} \hat{L}(
\hat{\rho}_R+\hat{\rho}_{NR })=0, \ee \be \frac{\partial
\hat{\rho}_{NR}}{\partial t}+i(1-\hat{D}) \hat{L}(
\hat{\rho}_R+\hat{\rho}_{NR })=0. \ee

For the purpose to obtain closed equation for $ \hat{\rho}_R $ we
exclude from the equation (104) $\hat{\rho}_{NR}$. As a result we
get \be \frac{\partial \hat{\rho}_R(t)}{\partial t}+i\hat{D}
\hat{L}\hat{\rho}_R(t)
+\int^{t}_{t_o}K(t-t_1)\hat{\rho}_R(t_1)dt_1=- i\hat{D}
\hat{L}exp[-i(t-t_o)(1-\hat {D})\hat{L}]\hat{\rho}_{NR}(t_o), \ee
\be K(t-t_1)= \hat{D} \hat{L}exp[-i(t-t_1)(1-\hat
{D})\hat{L}](1-\hat {D})\hat{L}. \ee This equation is valid for
$t=\sqrt{2}/\sigma>\tau_c$. Evidently $\hat{\rho}_R(t)$ is
expressed by way of values of $\hat{\rho}_R(t_1)$ taken for the
time interval $t_o<t_1\leq t$, and additionally through the value
of  $\hat{\rho}_{NR}(t_o)$. If in the initial moment of time
$t=t_o$ the system is in a pure quantum-mechanical state, then
$\hat{\rho}_{NR}(0)\neq 0$.

In this case solving the equation (107) is problematical. However
let us recollect that for $t_o>\sqrt{2}/\sigma$ system is already
in a mixed state. Therefore equation (107) takes more simple form
$(\hat{\rho}_{NR}(0)= 0)$ \be \frac{\partial
\hat{\rho}_R(t)}{\partial t}+i\hat{D} \hat{L}\hat{\rho}_R(t)
+\int^t_{t_o}K(t-t_1)\hat{\rho}_R(t_1)dt_1=0. \ee For solving
equation (108) we shall use a method of superoperators
\cite{Fujita}. But before performing this, we should note, that we
have obtained a closed equation for relevant part of statistical
operator. We were able to come to this only because when
$t>t_o=\sqrt{2}/\sigma$ the system is in the mixed state and all
nondiagonal matrix elements of the density matrix are equal to
zero.

Further when studying the evolution of the system we shall
consider as the origin of time the moment of the formation of the
mixed state in the system. This corresponds to a formal transition
to limit $t_o \rightarrow-\infty $. Further for simplification of
(108) we shall use Abel theorem \be \lim_{T\rightarrow\infty}
\frac{1}{T}\int^{0}_{T}f(t)dt=f(0)-\lim_{\varepsilon\rightarrow+0}
\int^{0}_{-\infty}e^{\varepsilon(t^{\prime})}\frac{d}{dt^{\prime}}f^{\prime}(t^{\prime})dt^{\prime}.
\ee Taking (109) into account, equation (108) will have the
following form: \be \frac{\partial \hat{\rho}_R(t)}{\partial
t}+i\hat D \hat{L}\hat{\rho}_R(t)
=\lim_{\varepsilon\rightarrow+\infty}
\int^{t}_{-\infty}e^{\varepsilon(t^{\prime}-t)}K(t-t^{\prime})\hat{\rho}_R(t^{\prime})dt^{\prime}.
\ee According to the method of superoperators \cite{Nakajima}, the
correspondence of one operator to another may be considered as
representation. The operator   in this case will be represented by
a matrix element with two indices, while the linear product of
operators is a matrix with four indices, i.e. a superoperator
\cite{Nakajima}. A concrete example is the projection of $\hat{D}$
operator on diagonal elements. We should notice, that in our case,
the projection operator is some definite physical procedure of
averaging matrix elements of the density matrix over Gaussian
chaotic ensemble.

From the relation $\hat{D}\rho_{nm}(t)= \rho_{nm}(t)\delta _{mn}$
we come to the following representation of $\hat{D}$ superoperator
$\hat{D}_{nmn^{\prime}m^{\prime}}=\delta _{nn^{\prime}}\delta
_{mm^{\prime}}\delta _{nn}$ so that \be
\sum_{m^{\prime}n^{\prime}}\hat{D}_{mnm^{\prime}n^{\prime}}\rho_{m^{\prime}n^{\prime}}=\rho_{nm}\delta
_{nm}. \ee Taking (111) into account, (110) is \be \frac{d
\rho_{nn}(t)}{dt}+i\hat{D} (\hat{L}\hat{\rho}_R)_{nn}= -
\int^{0}_{-\infty}
dt^{\prime}[K(-t^{\prime})\hat{\rho}_R(t+t^{\prime})]_{nn}
e^{\varepsilon t^{\prime}}. \ee In expression (112) and further
for short we shall omit $R$ index for diagonal matrix elements of
the operator $\hat{\rho}$. Considering the relation \be [\hat
{L},\hat{\rho}_R]_{nn}=\sum_{a} L_{nnaa}\rho_{aa}=0 \ee and
representing Liouville operator as $L=L_o+L^{\prime}$ in
compliance with (113) we get \be \frac{d
\rho_{nn}(t)}{dt}=-\int^{t}_{-\infty}dt_1 e^{\varepsilon(t-
t_1)}\sum_{n} K_{nnmm}(t-t_1)\rho_{mn}(t_1), \ee and \be
K_{nnmm}(t)=[L^{\prime}e^{-it(1-D)L}(1-D)L^{\prime}]_{nnmm}. \ee
From (115), (114) and from representation of Liouville operator in
the form $L=L_o+L^{\prime}$, one can see, that kernel $
K_{nnmm}(t)$ is at least of second order by $\Delta x_o$. Next it
is easy to check correctness of the expression \be \sum_{m}
L^{\prime}_{abmm}=\sum_{m} (V_{am}\delta _{bm}- V_{mb}\delta
_{am})=V_{ab}-V_{ab}=0, \ee for Liouville superoperator \be
L_{mnm^{\prime}n^{\prime}}=( H_{mm^{\prime}}\delta _{nn^{\prime}}-
H_{nn^{\prime}}\delta  _{mm^{\prime}}). \ee The relation (116) in
its turn leads to the rule of sums $\sum_{m}K_{mmnn}=0$. Taking
the symmetries $L_{abcd}=L_{cdab},~ D_{abcd}=D_{cdab}$ into
account, from (76) we get  \cite{Skrinnikov} \be \frac{d
\rho_{nn}(t)}{dt}=-\int^{t}_{\infty}dt_1 e^{\varepsilon (t-
t_1)}\sum_{m\neq n} [K_{nnmm}(t-t_1)\rho_{mm}(t_1)-
K_{mmnn}(t-t_1)\rho_{nn}(t_1)]. \ee Next we shall make the
following approximations. With accuracy up to the value of
$(\Delta x_o)^2$ order in the exponent in expression (115), we
shall replace the complete Liouville operator
$\hat{L}=\hat{L}_0+\hat{L}^{\prime}$  with $\hat{L}_0$. With the
same precision we may set $\rho_{nn}(t-t^{\prime})=\rho_{nn}(t)$.
As a result from (118) we get \be \frac{d \rho_{nn}(t)}{dt}=\sum
_{m\neq n}[W_{nm}\rho_{mm}(t)-W_{mn}\rho_{nn}(t)], \ee where \be
W_{nm}=-\int^{o}_{-\infty}dt^{\prime} e^{\varepsilon
t^\prime}[\hat{L}^{\prime}e^{it(1-\hat{D})\hat{L}_0}(1-\hat{D})\hat{L}^{\prime}]_{nnmm}.
\ee Since \be (\hat{D}\hat{L}_0)_{abcd}=\delta _{ab}[\Phi_a\delta
_{ac}\delta _{bd}-\Phi_b\delta _{bd}\delta _{ac}]=0, \ee in
expression (120) the operator $\hat{D}$ in the argument of
exponential function may be omitted. Taking into account the time
dependence of the operator $x(t)=x_o\Delta x_o
f(t),~f(t)\sum\limits^\infty_{\nu=-\infty}e^{i\nu\Omega
t},~\Omega=2\pi/T_o,$ for nondiagonal matrix elements we get
\cite{Skrinnikov}: \be \frac{d \rho_{nn}(t)}{dt}=\sum_{m\neq n}
[W_{nm}\rho_{mm}(t)-W_{mn}\rho_{nn}(t)], \ee where \be
W_{nm}=\frac{\pi}{2}|V_{nm}|^2\sum^{\infty}_{\nu=-\infty}\delta
(E_{nm}-\nu\Omega), \ee is the transition amplitude between the
eigenstates of the Hamiltonian
$\hat{H}_o,~E_{nm}=\varphi_n-\varphi_m,~V_{nm}$ is the matrix
element of the operator $\hat{V}_o=\Delta x_o Q^2_1Q^2_2$ in the
basic of eigenfunctions of the Hamiltonian
$\hat{H}_o,~V_{nm}=\langle\psi_n|\hat{V}_o|\psi_m\rangle$.

Equation (122) describes a nonreversible evolution of the system
from nonstationary state to the stationary state defined by the
principle of detail equilibrium. To prove irreversibility of the
process let us consider time dependence of nonequilibrium entropy
\cite{Fujita} \be S(t)=-K_B\sum_{n} \rho_{nn}(t)ln (\rho_{nn}(t)),
\ee where $K_B$ in the Boltzmann constant. Taking into account
$\sum_{n}\rho_{nn}(t)=1$, from (124) we get
$$\frac{dS(t)}{dt}=-K_B\sum_{n}\sum_{m} W_{nm}[\rho_{mm}(t)- \rho_{nn}(t)]ln(\rho_{nn}(t))-$$
\be -K_B\sum_{n}\frac{\partial\rho_{nn}(t)}{\partial
t}=\frac{1}{2} K_B\sum_{n}\sum_{m} W_{nm}[\rho_{nn}(t)-
\rho_{mm}(t)] [ln(\rho_{nn}(t))-ln(\rho_{mm}(t))]. \ee Due to the
property of logarithmic function \be (\rho_{nn}(t)-
\rho_{mm}(t))(ln(\rho_{nn}(t))-ln(\rho_{mm}(t)))\geq 0, \ee we see
that \be \frac{dS}{dt}\geq0 \ee This testifies the growth of
entropy during the evolution process.

More exact estimation of entropy growth may be obtained from the
principle of detail equilibrium $W_{nm}\rho_{mm}=
W_{mn}\rho_{nn}$. Taking into account that in our case
$W_{nm}=W_{mn}$, for the entropy growth we get \be \Delta S=
S(t\gg\sqrt{2}/\sigma)-S(t=0)=K_BlnN, \ee where $N$ is the number
of levels included into the process.

Thus, we can conclude, that the complex structure of energy
spectrum of quantum chaotic system leads to the fact that after
averaging over small spread in values of parameters there occurs
formation of a mixed sate in the system.

\section{Conclusion}

The traditional notion of an area, where the laws of statistical
physics are effective, consists of the assumption that the number
of interacting particles is sufficiently large. However, a lot of
examples of systems with a small number of degrees of freedom,
where chaotic motions occur, had become known by the end of the
last century. A new stage in the development of notions about
chaos and its origin appeared in the last two decades of the last
century. It turned out that the classical Hamiltonian system may
experience a special kind of instability. Because of this
instability various dynamic characteristics of the system randomly
change with time. Such a property of the system that performs
random motion is called dynamic stochasticity. It is well known
that the appearance of non-reversibility in classical chaotic
systems is connected with the local instability of phase
trajectories relatively to a small change of initial conditions
and parameters of the system. Classical chaotic systems reveal an
exponential sensitivity to these changes. This leads to an
exponential growth of the initial error with time, and as the
result after the statistical averaging over this error, the
dynamics of the system becomes non-reversible. In spite of this,
the question about the origin of non-reversibility in quantum case
remains actual. The point is that the classical notion of
instability of phase trajectories loses its sense during quantum
consideration. Therefore more detailed analyzes of the possible
mechanisms of irreversibility in chaotic quantum-mechanical
systems is needed. More over even the fact of possibility of
emergence of irreversibility in chaotic quantum-mechanical systems
is not evident. Quantum dynamics is unitary and therefore
reversible. But on the other hand, without assumption of small
initial dispersion of the system's parameters, classical chaotic
dynamics also will be reversible. Note that while studying
classical chaos, one usually examines the stability of the system
relatively to a small change of initial conditions and system
parameters. A small initial error of these parameters always
exists and remains unavoidable (one may measure the parameters of
the system at a very high precision, but even in this case there
is still a small error, the removal of which, i.e. the measurement
at absolute precision, is impossible). So not the existence of the
unavoidable error is fundamental, but what kind of influence it
brings over the system dynamics. It is well known that in case of
regular systems such an influence is negligible, but if the system
is chaotic, the effect of it increases exponentially. In this
review paper we touched upon results obtained in our previous
works. We studied paradigmatic model for quantum chaos, time
dependent universal Hamiltonian (Hamiltonian of mathematical
pendulum). Main result obtained in those works is that reason of
irreversibility in time dependent quantum dynamics of the chaotic
quantum-mechanical systems, is a very specific spectral
characteristic of the system. For universal Hamiltonian and
corresponding to it Mathieu-Schrodinger equation above mentioned
means very specific dependence of the energy eigenvalues on the
values of potential barrier (see. Fig.1). In addition, transition
form pure quantum-mechanical state to the mixed one, was found to
be quantum analog of the classical chaotic layer, characterized
classical chaotic motion near the separatrix. Results obtained are
important not only for better understanding of physics of quantum
chaos, but for practical implementation to organic molecules,
quantum computing and so on.

\textbf{Acknowledgments}

The designated project has been fulfilled by financial support
from the Georgian National Foundation (grants: GNSF/STO 7/4-197,
GNSF/STO 7/4-179). The financial support of Deutsche
Forschungsgemeinschaft SFB 484 under contract number 21095192(EC
94/5-1) is gratefully acknowledged by L. Chotorlishvili.


\begin{thebibliography}{40}
%
\bibitem{Sagdeev} R.Z. Sagdeev, D.A. Usikov, and G.M. Zaslavsky, Nonlinear Physics,Harwood Academic, New York
(1988).
%
\bibitem{Lichtenberg} A.J. Lichtenberg and M.A. Lieberman, Regular and Stochastic
Motion, Springer-Verlag, New York (1983).
%
\bibitem{Alligood} K.T. Alligood, T.D. Sauer, and J.A. York, Chaos: An Introduction to Dynamical Systems, Springer, New York
(1996).
%
\bibitem{Honggi} M.G. Grifoni and P. Hanggi, Phys.Rep. 304, 229 (1998).
%
\bibitem{Peres} A. Peres, Phys. Rev.A \textbf{30}, 1610 (1984).
%
\bibitem{Jacquod} P. Jacquod, C. Petitjean arXi 0806.0987 (2008).
%
\bibitem{Haake} F. Haake, Quantum Signatures
of Chaos, Springer, Berlin (2001).
%
\bibitem{Stockman} H.J. Stockman,
Quantum Chaos, An Introduction, Cambridge Univ. Press, Cambridge
(1993).
%
\bibitem{Ugulava} A. Ugulava, L.Chotorlishvili, and K.Nickoladze, Phys.Rev. E
\textbf{68}, 026216 (2003).
%
\bibitem{Chotorlishvili} A. Ugulava, L.Chotorlishvili, and K.Nickoladze, Phys.Rev. E
\textbf{70}, 026219 (2004).
%
\bibitem{Nickoladze} A. Ugulava, L.Chotorlishvili, and K.Nickoladze, Phys.Rev. E
\textbf{71}, 056211 (2005).
%
\bibitem{Gvarjaladze} A. Ugulava, L.Chotorlishvili, T. Gvarjaladze, and S.Chkhaidze,
Mod.Phys. Lett. B, \textbf{21}, 415 (2007).
%
\bibitem{Berman} G.P. Berman, G.M. Zaslavsky,
Phys. Lett. A, \textbf{61}, 295 (1977).
%
\bibitem{Janke-Emde-Losh} Janke-Emde-Losh, Tafeln Hoherer
Funktionen, Stutgart, 1960.
%
\bibitem{Bateman} H. Bateman, A. Erdelyi, Higher Transcedental
Functions, New-York, Toronto, London MC Graw-Hill Book Company
INC, 1955.
%
\bibitem{Hamermesh}  M. Hamermesh, Group Theory and its Application to Physical Problems,
1954.
%
\bibitem{Landau} L.D. Landau and E.M. Lifshitz, Quantum Mechanics, Non-relativistic Theory, Pergamon, Oxford
(1977).
%
\bibitem{Herzberg}  G. Herzberg, Infrared and Raman
Spectra of Polyatomic molecules, New York (1945).
%
\bibitem{Flygare} W.H. Flygare, Molecular Stucture and Dinamics (Prentice-Hall, Inc.,Englewood Cliffs, New Jersey,1978).
%
\bibitem{Essers} R. Essers, J. Tennyson and
P.E.S.Wormer, Chemical Physics Letters, \textbf{89}, 223 (1982).
%
\bibitem{Arranz} F.J. Arranz, F. Borondo, R.M. Benito, The Europian Physical Jornal D
\textbf{4}, 181 (1998).
%
\bibitem{Kaempffer} F.A. Kaempffer, Concepts in Quantum
Mechanics, 1965.
%
\bibitem{Bohm} D. Bohm, Quantum Theory (Prentice-Hall,
New-York, 1952).
%
\bibitem{Shimanouchi} T. Shimanouchi, Tables of Molecular
Vibrational Frequencies  Consolidated, National Bureau of
Standards, 1, 1-160 (1972).
%
\bibitem{Bloembergen} N. Bloembergen, E. Parcell and R. Pound, Phys.Rev.\textbf{73}, 679 (1948).
%
\bibitem{Gaspard} P. Gaspard, Chaos,
Scattering and Statistical Mechanics, Cambridge Univ. Press,
Cambridge (1998).
%
\bibitem{Hiller} M. Hiller, T.
Kottos, D. Cohen, T. Geisel, Phys.Rev. Lett. \textbf{92}, 010402
(2004).
%
\bibitem{Turchett} Q.A. Turchett et al. Phys. Rev. Lett. \textbf{7}, 4710 (1995).
%
\bibitem{Haug} F. Haug, M. Bienert, W. Schleich, T. Seligman, M. Raizen,
Phys. Rev. A, 71, 043803 (2005).
%
\bibitem{Prosen} T. Prosen, M. Znidaric, J.Phys. A, \textbf{35},1455 (2002).
%
\bibitem{Beneti} G. Beneti, G. Casati, Phys. Rev. E \textbf{65}, 066205 (2002).
%
\bibitem{Jalabert} R. Jalabert, H. Pastawski, Phys. Rev. Lett.
\textbf{86}, 2490 (2001).
%
\bibitem{Relano} A. Relano, J.M. Gomez, R.A. Molina, J.
Retamosa and E. Faleiro, Phys.Rev.Lett. \textbf{89}, 244102
(2002).
%
\bibitem{Retamosa} A. Relano, J. Retamosa, E. Faleiro, and
J.M. Gomez, Phys.Rev. E \textbf{72}, 066219 (2005).
%
\bibitem{Feynman} R.P. Feynman, Statistical Mechanics,
W.A. Benjaman Inc. Massachusetts (1972).
%
\bibitem{Schleich} W.P. Schleich, Quantum Optics in
Phase Space, Wiley-VCH, Berlin (2001).
%
\bibitem{Skrinnikov} L. Chotorlishvili, V. Skrinnikov, Phys.Lett.A, \textbf{372}, 761 (2008).
%
\bibitem{Kereselidze} A. Ugulava, L. Chotorlishvili, T. Kereselidze and V.
Skrinnikov,Mod. Phys. Lett. B\textbf{21}, 79 (2007).
%
\bibitem{Fujita} S. Fujita, Introduction to Non-Equilibrium Quantum Statistical Mechanics (W.B.Saunders
Company, Philadelphia-London, 1966).
%
\bibitem{Nakajima} S. Nakajima, Progr.Theor.Phys., \textbf{20}, 948 (1958).
%
\end{thebibliography}
\end{document}